\documentclass[twocolumn,preprintnumbers,floats,prd,amssymb,floatfix,nofootinbib,balancelastpage,superscriptaddress,amsmath]{revtex4-1}

\pdfoutput=1
\usepackage{graphicx,color,float}
\usepackage{hyperref}
\usepackage{multirow}
\usepackage{verbatim}
\usepackage{longtable}
 \usepackage{amsmath}
\usepackage{amsfonts}
\usepackage{amssymb}
\usepackage{rotating}

\usepackage[normalem]{ulem}

\def\gsim{\raise0.3ex\hbox{$\;>$\kern-0.75em\raise-1.1ex\hbox{$\sim\;$}}}
\def\lsim{\raise0.3ex\hbox{$\;<$\kern-0.75em\raise-1.1ex\hbox{$\sim\;$}}}

\begin{document}


\title{Time-delayed electrons from neutral currents at the LHC}
\renewcommand{\thefootnote}{\arabic{footnote}}

\author{Kingman Cheung}
\email{cheung@phys.nthu.edu.tw}
\affiliation{Department of Physics, National Tsing Hua University,	Hsinchu 300, Taiwan}
\affiliation{Center for Theory and Computation, National Tsing Hua University,	Hsinchu 300, Taiwan}
\affiliation{Division of Quantum Phases and Devices, School of Physics, Konkuk University, Seoul 143-701, Republic of Korea}

\author{Kechen Wang}
\email{kechen.wang@whut.edu.cn}
\affiliation{Department of Physics, School of Science, Wuhan University of Technology, 430070 Wuhan, Hubei, China}

\author{Zeren Simon Wang}
\email{wzs@mx.nthu.edu.tw}
\affiliation{Department of Physics, National Tsing Hua University,	Hsinchu 300, Taiwan}
\affiliation{Center for Theory and Computation, National Tsing Hua University,	Hsinchu 300, Taiwan}

\date{\today}

\begin{abstract}
 
 We investigate long-lived particles (LLPs) produced in pair from neutral currents and decaying into a displaced electron plus two jets at the LHC, utilizing the proposed minimum ionizing particle timing detector at CMS.
 We study two benchmark models: the R-parity-violating supersymmetry with the lightest neutralinos being the lightest supersymmetric particle and two different $U(1)$ extensions of the standard model with heavy neutral leptons (HNLs).
 The light neutralinos are produced from the standard model $Z$-boson decays via small Higgsino components, and the HNLs arise from decays of a heavy gauge boson, $Z'$.
 By simulating the signal processes at the HL-LHC with the  center-of-mass energy $\sqrt{s}=$ 14 TeV and integrated luminosity of 3 ab$^{-1}$, our analyses indicate that the search strategy based on a timing trigger and the final state kinematics has the potential to probe the parameter space that is complementary to other traditional LLP search strategies such as
 	those based on the displaced vertex. 
  
\end{abstract}


\vskip10mm

\maketitle
\flushbottom

\section{Introduction}\label{sec:intro}

In recent years there is a surge of interest in long-lived particles (LLPs) featuring new physics beyond the Standard Model (BSM).
This is not only due to the non-observation of new particles at the Large Hadron Collider (LHC) so far, but also to the fact that LLPs are widely predicted in many BSM models (see Refs.~\cite{Curtin:2018mvb,Lee:2018pag,Alimena:2019zri} for reviews of LLPs searches and models).
LLPs can be charged or neutral, and the latter is usually more challenging to search for in experiments.
For instance, a class of ``portal-physics'' models proposed to explain the dark matter or the non-vanishing neutrino masses often predict existence of
long-lived neutral light mediators of different spins, such as heavy neutral leptons (HNLs) that mix with the active neutrinos \cite{Gorbunov:2007ak,Atre:2009rg,Drewes:2013gca,Drewes:2015iva,Deppisch:2015qwa} and dark photons that arise from kinetic mixings in the minimal $U(1)$ extension of the Standard Model (SM) \cite{Okun:1982xi,Galison:1983pa,Holdom:1985ag,Boehm:2003hm,Pospelov:2008zw}.
Such LLPs can be produced at colliders and may have eluded experimental searches, simply because of our previous choices of search strategies aimed
mainly at new heavy particles decaying promptly.
Indeed, the ATLAS and CMS collaborations have performed various
LLP searches with distinct collider signatures, including disappearing
tracks \cite{Sirunyan:2018ldc,Sirunyan:2020pjd,ATLAS:2021ttq},
displaced leptons \cite{Khachatryan:2014mea,CMS:2016isf},
and heavy charged particles \cite{Aaboud:2019trc}.
Moreover, a number of far detectors
designed specifically for LLP searches have been proposed to be
installed with a distance of $5-500\text{ m}$
from different interaction points (IPs) of the LHC.
These include for instance FASER \cite{Feng:2017uoz,Ariga:2018uku},
MATHUSLA \cite{Curtin:2018mvb,Chou:2016lxi,Alpigiani:2020tva}, and
MoEDAL-MAPP \cite{Pinfold:2019nqj,Pinfold:2019zwp}.
In particular, FASER and MoEDAL-MAPP1 have been approved for operation
during the upcoming LHC Run-3, in the hope of observing displaced-decay events in the vicinity of the ATLAS and LHCb IPs, respectively.
These far-detector experiments mainly search for signatures such as displaced tracks or vertices.
Beyond these programs, future lepton colliders have also received
substantial attention for LLP searches at Higgs factories and muon colliders (see e.g. Refs.~\cite{Alipour-Fard:2018lsf,Cheung:2019qdr,Capdevilla:2021fmj})

A novel general strategy to search for LLPs was proposed in
Ref.~\cite{Liu:2018wte}, focusing on the time-delay feature of
the LLP decay products, which arises mainly from the nonrelativistic
{speed of not-so-light LLPs.
Precision timing upgrades have been proposed at various LHC experiments,
including ATLAS \cite{Allaire:2018bof,Allaire:2019ioj},
CMS \cite{CERN-LHCC-2017-027}, and LHCb \cite{Aaij:2244311}.
In particular, the future MIP (minimum ionizing particle) timing detector (MTD) at the CMS experiment is expected to have a time resolution of 30 picoseconds (ps)
for charged particles in the high-luminosity LHC (HL-LHC) era, and we will
focus on this setup in this study.
Such precision timing upgrades are originally purposed primarily for
reduction of pileup foreseen with the upcoming high-luminosity collisions.
However, incidentally they would also allow for discriminating the
LLP signatures from SM background, enhancing the sensitivities to
LLPs \cite{Liu:2018wte}.

Such enhancement brought by using the timing information was exemplified in Ref.~\cite{Liu:2018wte} for two benchmark models: SM Higgs decay to a pair of long-lived glueballs, and a long-lived neutralino in the gauge mediated supersymmetry breaking scenario.
Following that, Refs.~\cite{Du:2019mlc,Mason:2019okp} investigated
respectively long-lived dark photons in a model with two extra $U(1)$
gauge bosons, and HNLs from the SM Higgs decays in an effective model.
In particular, the results from the latter can be re-interpreted in terms of the $U(1)_{B-L}$ model which is considered here.
Additionally, Refs.~\cite{Kang:2019ukr,Cerri:2018rkm} proposed to use
the precision timing information for the determination of the LLP mass and
the identification of charged LLPs, respectively.
All these works have demonstrated the huge potential of the timing detector
for LLP searches in general, as well as its complementarity to the existing LHC searches in the studied scenarios.

In this work we move beyond to investigate two electrically neutral LLP candidates that are pairly produced in the same topological process at the LHC and also decay to almost identical final-state particles.
The first theoretical scenario is R-parity-violating (RPV) supersymmetry (SUSY) (see Refs.~\cite{Dreiner:1997uz,Barbier:2004ez,Mohapatra:2015fua} for reviews)
with the lightest neutralinos produced from the
SM $Z$-boson decays.
The lightest neutralino is dominantly bino-like with small Higgsino components coupled to the $Z$-boson.
The lightest neutralino may decay via an RPV coupling into leptons and jets.
The other physics scenario involves two slightly different $U(1)$ extensions of
the SM, predicting a new gauge boson, $Z'$, and three HNLs.
The $Z'$ boson can be produced directly on-shell from
proton-proton collisions and decays to a pair of HNLs which can
further decay to leptons and jets.
Both channels naively lack associated charged prompt objects
and entail a pair of charge-neutral LLPs, rendering the usual triggering and tracking difficult.
It presents a major challenge to experiments.
However, we will show in this paper the advantages of using
the timing information for this type of signal channels.

In Sec.~\ref{sec:models} we introduce the two models we consider.
Sec.~\ref{sec:search} details the CMS timing detector and our search
strategy, and discusses briefly the background estimate.
In Sec.~\ref{sec:numerical} we explain the simulation procedure and
present the numerical results.
We conclude the paper and provide an outlook in
Sec.~\ref{sec:conclusions}.

\section{Models}\label{sec:models}

In this work we focus on two benchmark models that would lead to LLPs pairly produced from neutral currents with very similar signatures.
Both the lightest neutralino and the HNLs become
long-lived because of light mass or feeble couplings to the SM particles.
We describe these models and the relevant constraints in detail in this section.

\subsection{Light neutralinos and the R-parity-violating supersymmetry}\label{subsec:model1}

One of the major deficiencies of the SM is the hierarchy problem \cite{Gildener:1976ai,Veltman:1980mj}.
A possible theoretical solution to this problem is to invoke a new symmetry between the fermions and bosons, known as the supersymmetry \cite{Nilles:1983ge,Martin:1997ns}.
The minimal realization of the theory with matter contents is known as the Minimal Supersymmetric Standard Model (MSSM).
In the MSSM, a priori a so-called ``R-parity'' is assumed as conserved, so as to attain proton stability at the renormalizable level.
It also has the consequence that the lightest supersymmetric particle (LSP) has to be stable and can serve as a dark matter candidate.
However, it is still legitimate to consider R-parity-violating SUSY (RPV-SUSY).
If the R-parity is broken, the MSSM superpotential is extended by the following operators:
\begin{eqnarray}
 W_\text{RPV}& =& \epsilon_i L_i \cdot H_u + \frac{1}{2}\lambda_{ijk} L_i \cdot L_j \bar{E}_k + \lambda'_{ijk} L_i \cdot Q_j \bar{D}_k \nonumber \\
   &&+ \frac{1}{2}\lambda^{''}_{ijk} \bar{U}_i \bar{D}_j \bar{D}_k,\label{eqn:RPVsuperpotential}
\end{eqnarray}
where the $i,j,k$ are generation indices.
These operators imply much richer phenomenological discussion both
at colliders and with low-energy observables.
Allowing all the operators to be nonvanishing would lead to a too fast proton decay rate, unless the RPV couplings are extremely tiny.
However, it is possible to consider a model where only certain sets of the operators in Eq.~\eqref{eqn:RPVsuperpotential} are present.
For example, if we invoke the baryon triality $B_3$ symmetry, the baryon-number-violating terms will vanish \cite{Ibanez:1991pr,Dreiner:2012ae}, while the others, which violate lepton number, would remain, thus the proton would not be predicted with a too fast decay rate \footnote{For a recent update on proton decays in the RPV-SUSY, see Ref.~\cite{Chamoun:2020aft}.}.
In this work, we choose to focus only on the operator $L_i \cdot Q_j \bar D_k$ with the generation indices $(i,j,k)=(1,1,2)$, as a benchmark scenario.

While the LHC has obtained stringent bounds on masses of squarks and gluinos \cite{Aaboud:2018doq,Sirunyan:2017nyt,Sirunyan:2019mbp,Sirunyan:2019ctn,Aad:2020nyj}, the mass of the lightest neutralino is relatively loosely constrained.
If the GUT (grand-unified-theory) relation $M_1\approxeq 1/2\, M_2$ on bino and wino masses and the dark matter assumption are dropped, the lightest neutralino can be as light as in the GeV scale or below \cite{Choudhury:1995pj,Choudhury:1999tn,Belanger:2002nr,Bottino:2002ry,Belanger:2003wb,Vasquez:2010ru,Calibbi:2013poa,Gogoladze:2002xp,Dreiner:2009ic}, at the same time in consistency with astrophysical and cosmological constraints \cite{Grifols:1988fw,Ellis:1988aa,Lau:1993vf,Dreiner:2003wh,Dreiner:2013tja,Profumo:2008yg,Dreiner:2011fp}.
Such light neutralinos have to be bino-like \cite{Gogoladze:2002xp,Dreiner:2009ic}, and should decay to avoid overclosing the Universe \cite{Hooper:2002nq,Bottino:2011xv,Belanger:2013pna,Bechtle:2015nua}.
In fact, one of the scenarios where the lightest neutralino can decay is exactly the RPV-SUSY.
With an RPV operator, e.g. $L_1 \cdot Q_1 \bar D_2$, the GeV-scale
neutralinos can decay via an off-shell sfermion to two quarks and a lepton,
at the parton level.
Further, given the required relatively small values of $\lambda'_{112}/m^2_{\tilde{f}}$, the lightest neutralino with mass $m_{\tilde{\chi}_1^0}\lesssim \mathcal{O}(10)$ GeV can become long-lived.
Once produced at a collider, it can travel a macroscopic distance before decaying, leading to spectacular collider signatures.

We extract the present bound on $\lambda'_{112}$ from Ref.~\cite{Allanach:1999ic}, which depends on the mass of the right-chiral strange squark, $m_{\tilde{s}_R}$:
\begin{eqnarray}
\lambda'_{112} < 0.21 \times \frac{m_{\tilde{s}_R}}{1\text{ TeV}}.\label{eqn:RPVbound}
\end{eqnarray}
For simplicity, for the rest of this work, we will assume degenerate sfermion masses.
Assuming the lightest neutralino is dominantly bino-like and is the LSP, we then estimate the proper decay length of the lightest neutralino, $c\tau_{\tilde{\chi}_1^0}$ and express it with the following formula:
\begin{eqnarray}
c\tau_{\tilde{\chi}_1^0} \simeq (2.8 \text{ m}) \Big(  \frac{m_{\tilde{f}}}{1\text{ TeV}} \Big)^4 \Big( \frac{10\text{ GeV}}{m_{\tilde{\chi}_1^0}} \Big)^5 \Big( \frac{0.01}{\lambda'_{112}}\label{eqn:ctaun1} \Big)^2,
\end{eqnarray}
We crosscheck Eq.~\eqref{eqn:ctaun1} by performing numerical evaluation with a UFO (Universal FeynRules Output) \cite{Degrande:2011ua} model file for the RPV-SUSY model \cite{RPVUFO} and MadGraph 2.7.3 \cite{Alwall:2014hca}, and find good agreement.
Compared to a similar expression, Eq.~(8) of Ref.~\cite{Wang:2019orr} for the same coupling $\lambda'_{112}$, Eq.~\ref{eqn:ctaun1} is slightly smaller by a factor of $\sim 1.144$.
In Fig.~\ref{fig:n1ctau} we show a plot of the lightest neutralino proper decay length versus mass, for different values of $\lambda'_{112}/m^2_{\tilde{f}}$.
\begin{figure}[t]
	\includegraphics[width=0.49\textwidth]{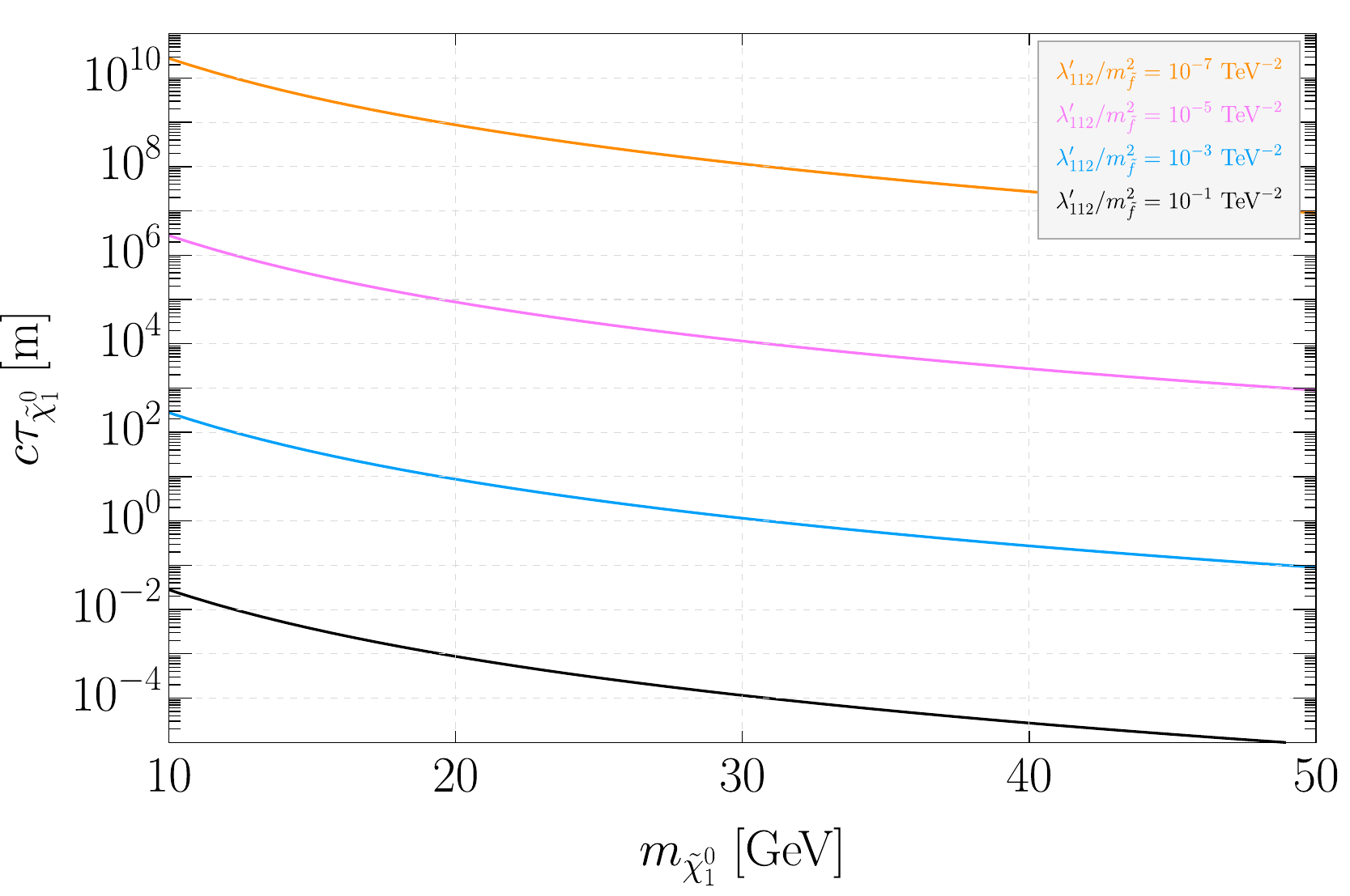}
	\caption{$c\tau_{\tilde{\chi}_1^0}$ vs. $m_{\tilde{\chi}_1^0}$, for various values of $\lambda'_{112}/m^2_{\tilde{f}}$.}
	\label{fig:n1ctau}
\end{figure}

At colliders and $B$-factories, such long-lived light neutralinos can be produced in different channels, including decays of $Z$-bosons, mesons, $\tau$ leptons, and squarks (see Refs.~\cite{Helo:2018qej,Dercks:2018eua,deVries:2015mfw,Dercks:2018wum,Wang:2019orr,Wang:2019xvx,Dreiner:2020qbi,Dey:2020juy,Gehrlein:2021hsk} for previous relevant studies).
In this work, we focus on the $s$-channel $Z$-boson decay into a pair of light neutralinos.
The SM $Z$-bosons do not couple to binos or winos, but the small components of Higgsinos in the GeV-scale neutralinos are sufficient to allow for large sensitivity reach at the LHC, by virtue of the large production rate of the $Z$-bosons.
With the current bounds on the Higgsino mass as well as the invisible decay width of the $Z$-boson, an upper bound of Br$(Z\to \tilde{\chi}_1^0 \tilde{\chi}_1^0)\sim 10^{-3}$ can be obtained \cite{Helo:2018qej}.
Thus in this work, we will treat Br$(Z\to \tilde{\chi}_1^0 \tilde{\chi}_1^0)$~\footnote{Here, Br$(Z\to \tilde{\chi}_1^0 \tilde{\chi}_1^0)$ is for negligible neutralino masses. In the numerical computation, we take into account the phase space effect, which becomes more important for $m_{\tilde{\chi}_1^0}\sim m_Z/2$.} as an independent parameter, together with $m_{\tilde{\chi}_1^0}$ and $\lambda'_{112}/m^2_{\tilde{f}}$.

\subsection{Heavy neutral leptons and $Z'$ in $U(1)_{B-L}$ and $U(1)_X$}\label{subsec:model2}

The observation of neutrino oscillation has firmly established the nonvanishing masses of the active neutrinos \cite{deSalas:2020pgw}.
As the SM explicitly entails massless neutrinos, BSM physics is required to invoke certain mechanisms for neutrino mass generation.
The most common way for this purpose is to introduce right-handed SM gauge singlets, which may, through a Yukawa-like term, couple to the active neutrino and the Higgs fields to induce a Dirac mass term.
For such singlets a Majorana mass term can also be written down in the Lagrangian.
The most classic model known as the Type-I seesaw mechanism \cite{Minkowski:1977sc,Yanagida:1979as,Gell-Mann:1979vob,Mohapatra:1979ia,Schechter:1980gr}, explains the small but nonzero active neutrino masses through mixings between active neutrinos and heavy (GUT-scale) right-handed neutrinos, often called sterile neutrinos.
There are also other similar models such as linear and inverse seesaw mechanisms \cite{Mohapatra:1986bd,Akhmedov:1995ip,Akhmedov:1995vm}, which allow for much lighter sterile neutrinos while keeping the active neutrino masses small.
The sterile neutrinos, once produced, can decay into a charged lepton via the mixing parameters.
If the mixing parameters are sufficiently small, these light sterile neutrinos can become long-lived.
Therefore, phenomenologically we can simply assume the sterile
neutrino masses, $m_N$, and the mixing parameters (squared) with the
active neutrinos, $V^2$, as two sets of independent parameters,
and call the sterile neutrinos as HNLs.

Beyond such minimal scenarios, HNLs are also predicted in a number of
more extended models including the left-right symmetric
model \cite{Mohapatra:1974gc,Pati:1974yy,Mohapatra:1980yp,Keung:1983uu},
leptoquark \cite{Dorsner:2016wpm}, and
a $Z'$ gauge boson \cite{Deppisch:2019kvs,Chiang:2019ajm}.
We focus on the latter case in this work.

We extend the SM gauge group by an $U(1)_X$, which is a linear combination of the SM $U(1)_Y$ and the $U(1)_{B-L}$ symmetries \cite{Mohapatra:1980qe}, often called as the ``non-exotic $U(1)_X$ model'' \cite{Appelquist:2002mw}.
\begin{table}[t]
	\begin{tabular}{c|ccc|c}
	         	& $SU(3)_c$  & $SU(2)_L$  & $U(1)_Y$ & $U(1)_X$ \\
		\hline
		$Q_L^i$ & 3          & 2& $\frac{1}{6}$ &  $\frac{1}{6}x_H+\frac{1}{3}x_\Phi$        \\
		$u_R^i$ & 3          & 1& $\frac{2}{3}$ & $\frac{2}{3}x_H+\frac{1}{3}x_\Phi$      \\
		$d_R^i$ & 3          & 1&$-\frac{1}{3}$ &    $-\frac{1}{3}x_H+\frac{1}{3}x_\Phi$       \\
		$L_L^i$ & 1          & 2&$-\frac{1}{2}$ &  $-\frac{1}{2}x_H- x_\Phi$        \\
		$e_R^i$ & 1          & 1& $-1$          &   $-x_H- x_\Phi$       \\
		$H$     & 1          & 2& $\frac{1}{2}$         &   $\frac{1}{2}x_H$       \\
		$N^i$   & 1          & 1&  $0$          &   $-x_\Phi$       \\
		$\Phi$  & 1          & 1&  $0$          &     $2x_\Phi$    
	\end{tabular}
	\caption{Field content and charge assignments in the non-exotic $U(1)_X$ model. $i=1,2,3$ is generation index. $\frac{1}{2}x_H$ and $2 x_\Phi$ are the $U(1)_X$ charges assigned for the SM Higgs boson $H$ and the new Higgs boson $\Phi$, with $x_H$ and $x_\Phi$ parameterizing the $U(1)_X$ charges of the fields in the model.}
	\label{tab:U1X}
\end{table}
The field content is listed in Table~\ref{tab:U1X}, together with the charge assignments.
As shown in Table~\ref{tab:U1X},
the $U(1)_X$ charges are controlled by two parameters: $x_H$ and $x_\Phi$.
Since the $U(1)_X$ charges are a linear combination of $U(1)_Y$ and $U(1)_{B-L}$, we fix $x_\Phi=1$ without loss of generality.
In the case of $x_H=0$ we recover the $U(1)_{B-L}$ model.
In the spectrum of the model, in addition to the SM particles, there are a new vector boson $Z'$, a new scalar particle  $\Phi$ which obtains a VEV (vacuum expectation value), $v_\Phi$, to break the $U(1)_X$ symmetry and mixes with the SM Higgs boson, and three right-handed neutrinos (or, equivalently, HNLs).
In particular, the three HNLs ensure that the model is free from gauge and gravitational anomalies \cite{Das:2016zue,Oda:2015gna}.

The SM Yukawa sector is now augmented with the following Dirac and Majorana terms:
\begin{eqnarray}
    \mathcal{L}_Y^{U(1)_X} = - Y_D \bar{L}_L \tilde{H} N -
    Y_N \Phi \overline{N^c} N + \text{h.c.},
\end{eqnarray}
where   $\tilde{H} = i \tau_2 H^*$ ($\tau_2$ is the second Pauli matrix),
  the superscript ``$c$'' denotes the charge conjugation, and
the generation indices are being suppressed.
After the electroweak and $U(1)_X$ symmetries are broken, the masses of the $Z'$ boson, and the Majorana and Dirac neutrinos are given by
\begin{eqnarray}
	m_{Z'} \simeq 2 g'_1 v_\Phi,	M_N = \frac{Y_N}{\sqrt{2}}v_\Phi, M_D = \frac{Y_D}{\sqrt{2}}v_H,
\end{eqnarray}
where $g'_1$ is the gauge coupling of $U(1)_X$ and $v_H=246$ GeV is
the SM Higgs VEV.
The LEP constraints require that $v_\Phi \gg v_H$
\cite{Carena:2004xs,Heeck:2014zfa}.
Diagonalization of the neutrino mass matrix then leads to the active neutrino mass matrix via the seesaw mechanism: $m_\nu \simeq -M_D M_N^{-1} M_D^T $ and the matrix of small mixing parameters between the active neutrinos and HNLs: $V_{lN} = M_D M_N^{-1}$.
For simplicity we assume there is only one HNL within the kinematically accessible range while the other two HNLs are much heavier, and we
consider this HNL is of Majorana nature and is mixed only with the SM
electron neutrino.

The $Z'$ boson can decay to a pair of SM fermions or HNLs.
The analytic expressions of these decay widths as
functions of $g'_1$, $U(1)_X$ charges as well as masses,
can be found in e.g. Refs.~\cite{Das:2017deo,Chiang:2019ajm}.
To maximize the decay branching ratio of $Z'$ into a pair of $N$'s,
we also consider a scenario with $x_H=-1.2$ \cite{Das:2017flq},
which we will call as $U(1)_X$ below, besides the minimal $U(1)_{B-L}$
case with $x_H=0$.

Previous searches for a $Z'$ boson at LEP \cite{LEP:2003aa},
Tevatron \cite{Carena:2004xs}, and LHC \cite{Amrith:2018yfb}
have set stringent bounds on $m_{Z'}$ and $g'_1$.
Here, we follow Ref.~\cite{Chiang:2019ajm} and choose to fix a benchmark
parameter point of $(m_{Z'},g'_1) = (6 \text{ TeV}, 0.8)$ throughout this paper,
which is allowed by the latest searches \cite{Sirunyan:2018exx,Aad:2019fac}.
\begin{figure}[t]
	\includegraphics[width=0.49\textwidth]{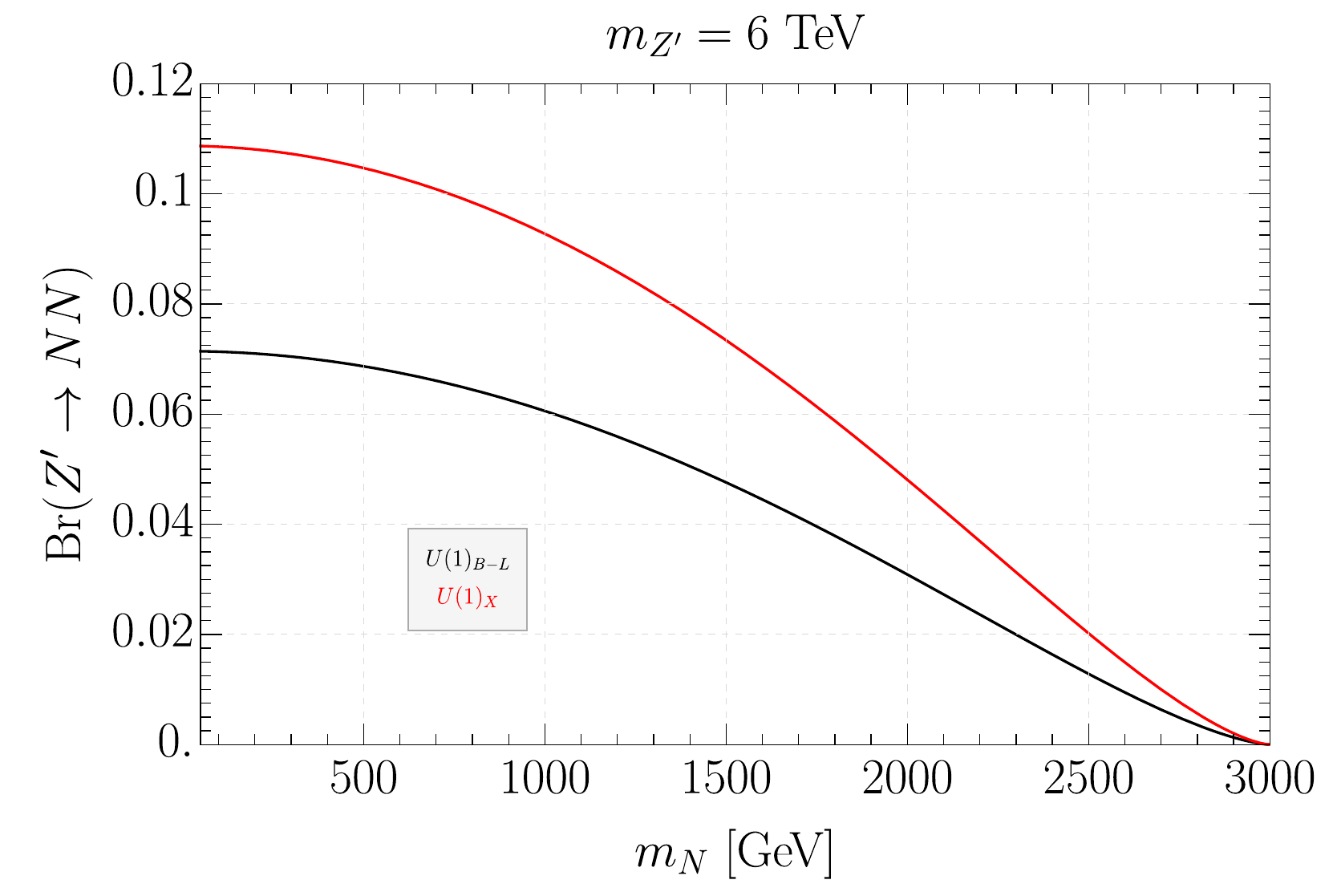}
	\caption{Br($Z'\to NN$) vs. $m_N$, for $m_{Z'}=6$ TeV.}
	\label{fig:BrZp2NN}
\end{figure}
In Fig.~\ref{fig:BrZp2NN} we show a plot in the plane
Br($Z'\to NN$) versus $m_N$ for $m_{Z'}=6$ TeV.

Thus, for the production of the HNLs, we study the process $pp\to Z' \to NN$, where the heavy $Z'$ boson with $m_{Z'}=6$ TeV is produced on-shell.
\begin{figure}[t]
	\includegraphics[width=0.49\textwidth]{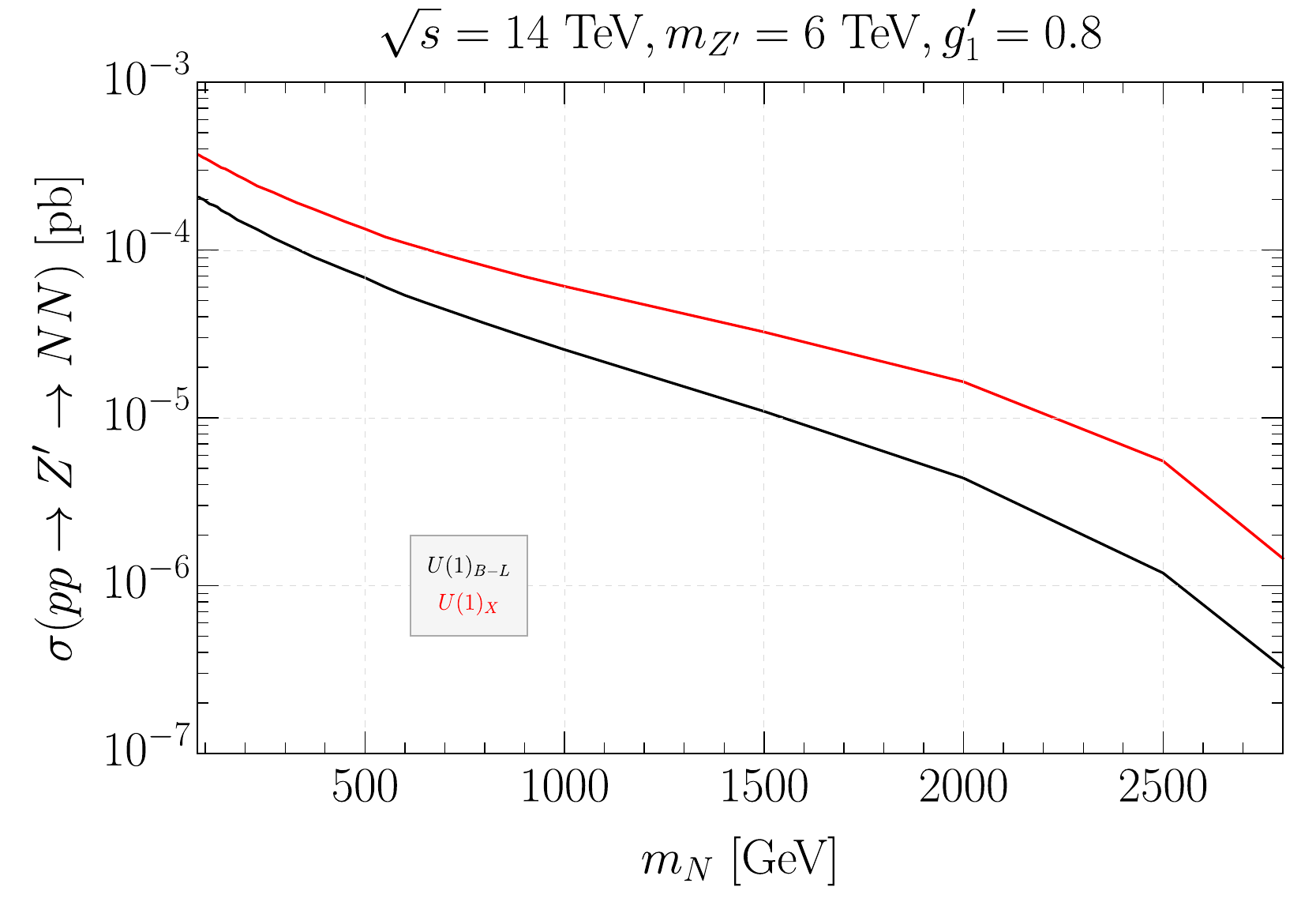}
	\caption{The \textit{inclusive} scattering cross section, $\sigma(pp\to Z' \to NN)$, in picobarn as a function of $m_N$, for $m_{Z'}=6$ TeV and $g'_1=0.8$.}
	\label{fig:XSpp2Zp2NN}
\end{figure}
We plot the \textit{inclusive} scattering cross section of $pp\to Z'\to NN$ in Fig.~\ref{fig:XSpp2Zp2NN} as a function of the HNL mass $m_N$, for $U(1)_{B-L}$ ($x_H=0$) and $U(1)_X$ ($x_H=-1.2$) models.
Note that we switch off the scalar mixing so that HNL pair production
can stem only from $Z'$ decays (except for the SM $Z$-decay which is doubly suppressed by the tiny mixing parameters squared and hence gives negligible contributions).

The decay of the HNL is mediated only via its mixing with
$\nu_e$, through charged-current ($W$-boson) and neutral-current ($Z$-boson)
interactions.
We compute the decay widths with analytic formulas given in
Refs.~\cite{Atre:2009rg,Bondarenko:2018ptm}.
\begin{figure}[t]
	\includegraphics[width=0.49\textwidth]{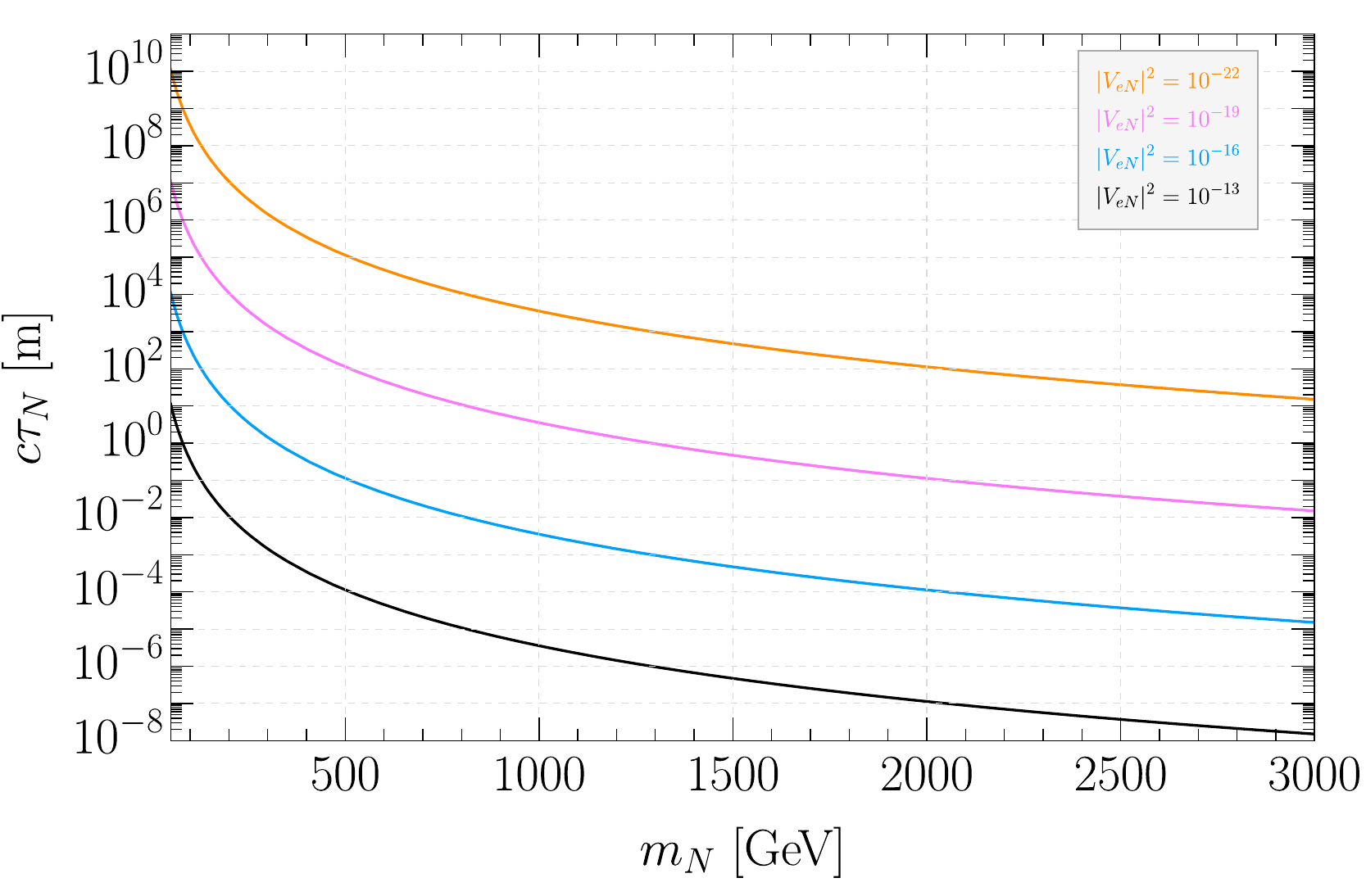}
	\caption{$c\tau_{N}$ vs. $m_{N}$, for different values of $|V_{eN}|^2$.}
	\label{fig:hnlctau}
\end{figure}
Figure~\ref{fig:hnlctau} contains a plot for the proper decay length of
the HNL, $c\tau_N$, as a function of $m_N$, for several choices of
the tiny $|V_{eN}|^2$ below $10^{-10}$.

Various experiments (colliders, beam dump experiments, neutrinoless double beta decay, etc.) have established constraints on the active-heavy neutrinos mixing parameters for different mass ranges.
As we will see later in the paper, our search strategy will be mainly sensitive to $m_N$ of $\mathcal{O}(100)$ GeV.
For this mass range, the strongest limits are currently only at the order of $10^{-3}$ for $|V_{eN}|^2$ \cite{Deppisch:2015qwa},
while the analyses here show that our search strategy could probe mixing parameters more than 10 orders of magnitude smaller.

\section{Timing detector and search strategy}\label{sec:search}

\begin{figure}[t]
	\includegraphics[width=0.49\textwidth]{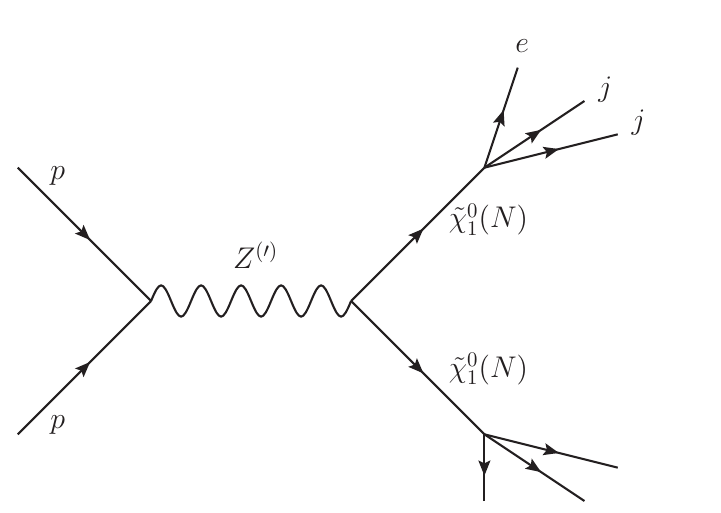}
	\caption{Feynman diagram of the signatures. We require at least one neutralino or HNL to decay into the $ejj$ final state, as well as one hard prompt ISR jet in the event analysis.}
	\label{fig:feynman}
\end{figure}
The signatures we focus on are shown in Fig.~\ref{fig:feynman}.
The $Z$ or $Z'$ boson is produced on-shell from proton-proton collisions and decays into a pair of light neutralinos or HNLs.
Then at least one of the two LLPs travels a macroscopic distance before decaying further into the $ejj$ final state.
Given the almost identical signatures for the two considered models, we apply the same search strategy for them.
The search is mainly based, as discussed earlier in Sec.~\ref{sec:intro}, on a timing trigger provided by the future upgrades in the CMS experiment.
In the rest of this section, we will introduce the CMS timing detector in more detail, explain the step-wise selection cuts, and briefly discuss the estimation of background sources.

\subsection{CMS minimum ionizing particle timing detector}\label{subsec:MTD}

As discussed in Sec.~\ref{sec:intro}, we
investigate the search potential of the CMS MTD for LLPs.
The timing layer is proposed to be installed between the inner tracker and the electromagnetic calorimeter, with a transverse distance of 1.17 m from the IP and a length of 6.08 m in the longitudinal direction.
With a high timing precision of 30 ps, it is possible to detect
signatures with a time delay in the nanosecond (ns) range.
We note that such a search would not require tracking information, potentially enhancing the sensitivities to parameter regions compared to the traditional displaced-object searches.

\subsection{Search strategy}\label{subsec:strategy}

We follow closely Refs.~\cite{Liu:2018wte,Du:2019mlc,Mason:2019okp} for pinning down the search strategy, which consists of a few event selections.
We start with a requirement on the transverse momentum and pseudorapidity of the leading electron in the event: $p_T^e > 20$ GeV and $|\eta^e| < 2.5$.
The LLP decay is then required to take place within the fiducial volume of the MTD, i.e. $200 < r < 1170$ mm and $|z| < 3040$ mm, for the transverse and longitudinal distances from the CMS IP.
The requirement of $r > 200$ mm ensures that the LLP decays outside the region of optimal tracking capabilities, so that the major SM background stems from trackless jets \cite{Liu:2018wte}. 
A signal event should also include at least one prompt ISR (initial-state-radiation) jet with $p_T^j > 30$ GeV.
We assume that it travels at the speed of light and use its location in pseudorapidity, its time of arrival, and the IP position, to timestamp the hard collision of the event.
This, together with the time of arrival and pseudorapidity of the leading electron, allows us to calculate the time delay of the electron, and we select only events with a time delay larger than 1 or 2 ns.

The time delay, $\Delta t$, is computed as follows,
\begin{eqnarray}
\Delta t  = t_{\text{arrival}}^{e} - t_{\text{prompt}}^{e},
\end{eqnarray}
where $t_{\text{arrival}}^{e}= \frac{l_{\text{LLP}}}{\beta_{\text{LLP}}}+l_e$ is the arrival time of the electron at the MTD, with $l_\text{LLP}$ ($l_e$) being the distance traveled by the LLP (displaced electron) before the LLP decays (the
electron hits the MTD), and $\beta_{\text{LLP}}$ being the speed of the
LLP, while $t_{\text{ prompt}}^{e}$ is the time when the electron would
arrive at the same position, \textit{had it been promptly produced at the IP and traveling with the speed of light}.

We would like to mention that naively this strategy can also be applied to the minimal scenario of HNLs, where the HNLs are produced directly from e.g. on-shell $W$ or $Z$ decays via the active-heavy neutrinos' mixings.
In particular, for the $W$ decay into, e.g., an electron and an HNL, no hard prompt ISR jet would be required, as the prompt electron can be used for timestamping the hard collision at the IP. 
However, as we show later in Sec.~\ref{sec:numerical} the timing-trigger-based strategy only receives high acceptance rates for long-lived HNLs with $m_N \gtrsim \mathcal{O}(10)$ GeV.
This would require a very small mixing parameter for keeping the HNL long-lived, which then in turn renders the production rates of the HNLs from $W$ or $Z$ decays too low to produce enough signal events.

\subsection{Background}\label{subsec:background}

Since our requirement on the transverse momentum of the leading electron and ISR jet mainly follows from Ref.~\cite{Mason:2019okp}, the estimate of background events should come to the same conclusion.
Therefore, given the unnecessity to repeat the computation, we will briefly explain the main background sources and summarize the final estimated numbers as given in Ref.~\cite{Mason:2019okp}.

Because of the finite timing resolution (30 ps), same-vertex (SV) hard collisions of jet and photon production may lead to a fake signal.
Considering the inclusive photon production, as well as jet production ($p_T^j > 30$ GeV) with a jet misidentified as a photon, the number of SV background events is found to be around $2\times 10^{11}$.
Using a Gaussian smearing with a time spread of 30 ps, a time-delay cut of $\Delta t > 1$ ns removes all of these background events.

A more important background source is the so-called pileup (PU) events.
These arise because in each bunch crossing there are multiple collisions taking place, which will be an important issue at the HL-LHC ($n_{\text{PU}} \sim 100$).
It is possible that besides the triggering hard collision, there is another pileup collision leading to a time-delay signal.
Taking the fraction of jets being trackless as $10^{-3}$, the total number of pileup events is estimated to be $10^7$.
Similar to the SV background events, a Gaussian smear of 190 ps spread gives the final background event number to be 0.7 and 0, for $\Delta t > 1$ ns and 2 ns, respectively.
Here, 190 ps is derived by the longitudinal spread of the bunch crossing.

Therefore, we conclude that with a cut of $\Delta t > 1$ ns, the background is essentially negligible.
Thus, when we discuss numerical results in the following section, we will focus on the $\Delta t > 1$ ns cut, while still examining the effect of a more strict time-delay cut.

\section{Numerical simulation and results}\label{sec:numerical}

We perform Monte-Carlo simulation in order to determine the exclusion limits for the two physics scenarios.

We simulate the SM $Z$-boson production and decay with Pythia8.243 \cite{Sjostrand:2014zea}.
These $Z$-bosons are set to decay exclusively into a pair of spin-1/2 fermions, in order to allow for obtaining the most statistics.
The vector and axial-vector couplings of the $Z$-boson to these new fermions are tuned to be the same as those for a $Z$-boson coupled to a pair of Higgsinos.
We also turn on the ISR effect in Pythia8, in order to obtain prompt ISR jets for timestamping the hard collision at the IP.
We then scan a two-dimensional grid, for 22 values of $m_{\tilde{\chi}_1^0}$ between 15 and 46 GeV, and 40 values of $\lambda'_{112}/m^2_{\tilde{f}}$ from $10^{-6}$ TeV$^{-2}$ to $9\times 10^{-3}$ TeV$^{-2}$ in logarithmic steps.
We simulate $10^8$ events at each grid point.

To generate the HNLs in the $U(1)_X$ and $U(1)_{B-L}$ models, we use the corresponding UFO model files as was used in Ref.~\cite{Chiang:2019ajm}, and generate parton-level events of $p p \to Z' \to N N$ at the leading order with MadGraph5 2.7.3 \cite{Alwall:2014hca}.
The simulated HNLs are then forced to decay exclusively to the $ejj$ final states with the tool of MadSpin \cite{Artoisenet:2012st} to ensure numerical stability for even very small decay widths.
The LHE output files from MadGraph5 are then processed to Pythia8 to include the ISR effects and provide the kinematics of the final state particles.
A 2D scan is then conducted.
For the $U(1)_{B-L}$ model, we scan $m_N$ from 300 to 700 GeV in intervals of 20 GeV, and $|V_{eN}|^2$ from $10^{-19}$ to $9\times 10^{-18}$ in 20 logarithmic steps, with one million events at each parameter point.
For the $U(1)_X$ model, given the larger scattering cross sections, we expect stronger exclusion limits, and hence simulate 200 thousand events for 35 HNL masses from 85 GeV up to 2400 GeV, and 70 values of $|V_{eN}|^2$ from $10^{-21}$ to $9\times 10^{-15}$, in logarithmic steps.

We express the total numbers of signal events for the two physics scenarios with the following formulas:
\begin{eqnarray}
N_s^{\tilde{\chi}_1^0} &=& N^Z  \cdot \text{Br}(Z\to \tilde{\chi}_1^0 \tilde{\chi}_1^0 ) \cdot \nonumber \\
	&&  \text{Br}(\tilde{\chi}_1^0\to e^- u \bar{s}\text{ or } e^+ \bar{u} s)   \cdot 2 \cdot \epsilon^{\tilde{\chi}_1^0},\label{eqn:signalnumber1} \\
N_s^N &=& \sigma^N  \cdot \mathcal{L}  \cdot \text{Br}(N\to ejj)   \cdot 2 \cdot \epsilon^N,\label{eqn:signalnumber2}
\end{eqnarray}
where $N^Z \simeq 1.9\times 10^{11}$ is the total number of $Z$-bosons resonantly produced with 3 ab$^{-1}$ integrated luminosity \cite{Helo:2018qej}, $\sigma^N$ is the \textit{inclusive} scattering cross section of $pp\to Z' \to NN$ calculated by MadGraph5, and $\epsilon^{\tilde{\chi}_1^0}$ and $\epsilon^{N}$ are the event acceptance rates including the requirement of one hard ISR prompt jet.
$\text{Br}(\tilde{\chi}_1^0\to e^- u \bar{s}\text{ or } e^+ \bar{u} s)=0.5$\footnote{The lightest neutralino can also decay to $\nu d \bar{s}$ or $\bar{\nu} s \bar{d}$ with a summed decay branching ratio of $50\%$, which does not include a time-delayed electron and is hence not considered as a signature in this work.}, and $\mathcal{L}=3$ ab$^{-1}$ labels the integrated luminosity at the HL-LHC.
Finally, the factor 2 that appears in both Eq.~\eqref{eqn:signalnumber1} and Eq.~\eqref{eqn:signalnumber2} accounts for the fact that in each signal event a pair of LLPs are produced and we require only one of them to decay into the specified final states.

As we discussed in Sec.~\ref{sec:search}, with the selection cuts we have chosen including the requirement of $\Delta t> 1$ or 2 ns, the background events can be considered to be negligible.
Therefore we will show 3-signal-event isocurves as 95\% confidence level (C.L.) exclusion limits in the numerical results.

\subsection{The light neutralino scenario}\label{subsec:results1}

\begin{figure}[t]
	\includegraphics[width=0.49\textwidth]{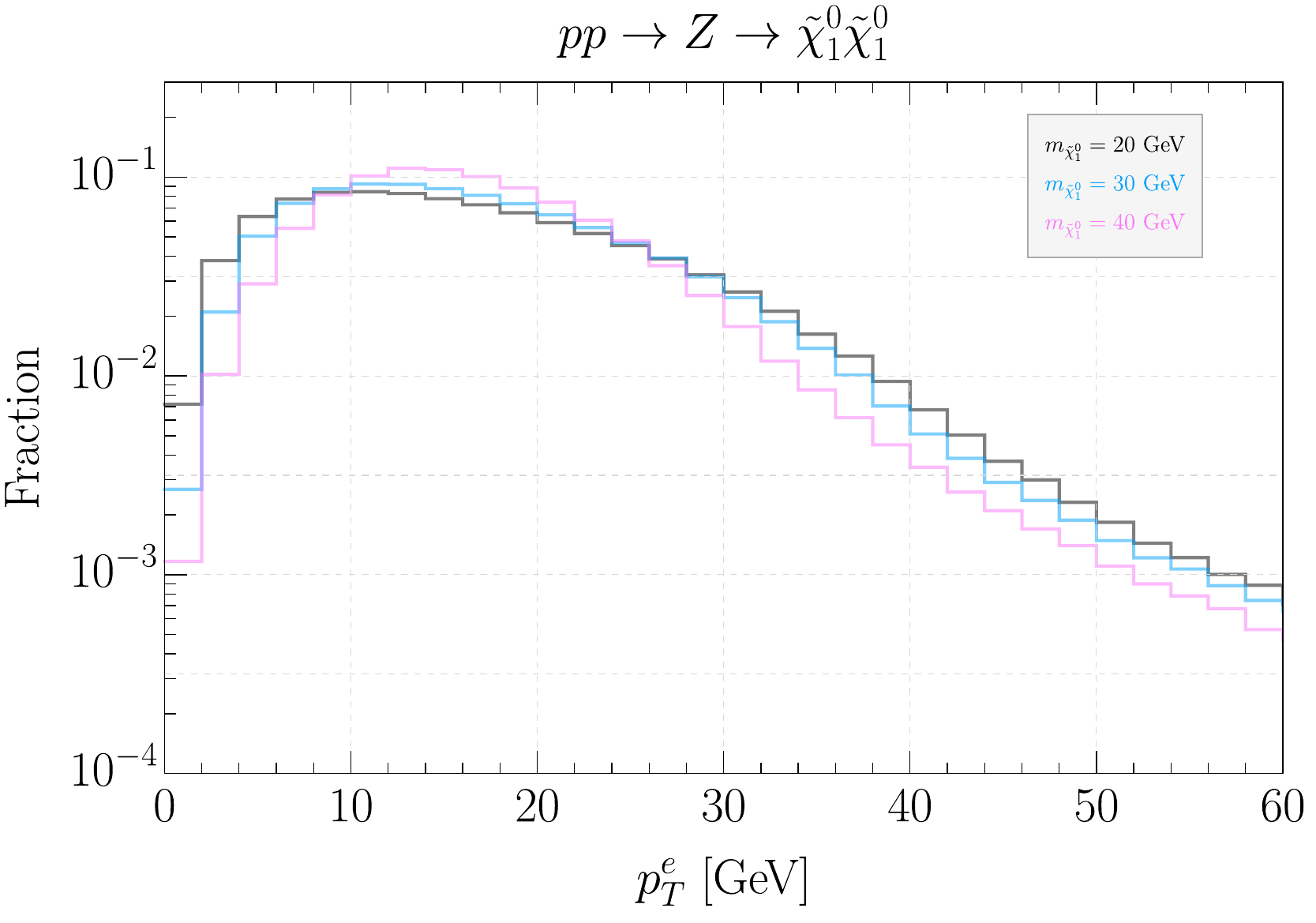}
	\includegraphics[width=0.49\textwidth]{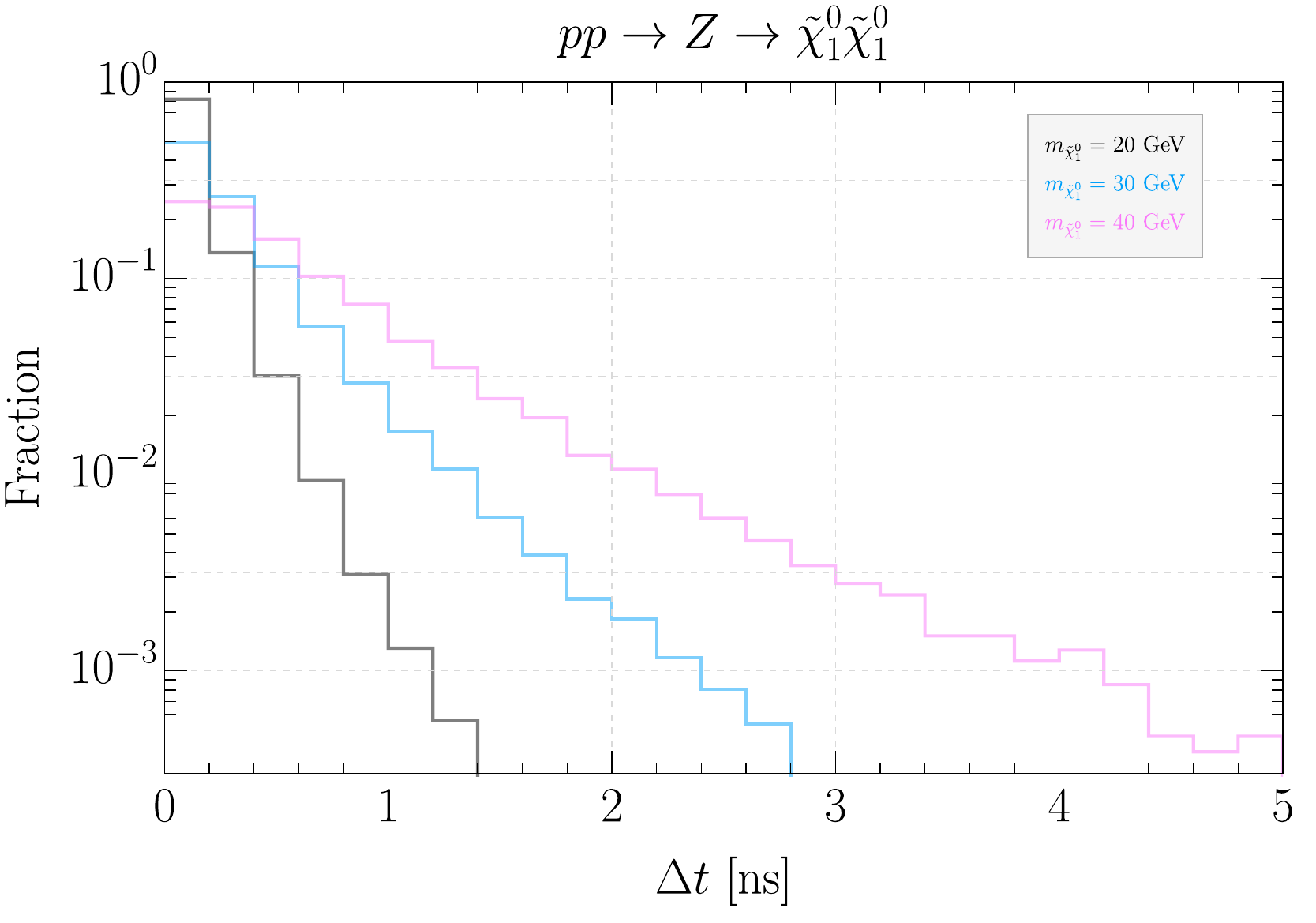}
	\caption{Distributions of $p_T^e$ and $\Delta t$ for light neutralinos. We fix $c\tau_{\tilde{\chi}_1^0}$ at 1 m.}
	\label{fig:distributions_n1}
\end{figure}
To present the numerical results for the light neutralino scenario,
we start with the kinematic distributions given in Fig.~\ref{fig:distributions_n1}.
The upper plot contains the distributions of the leading electron transverse momentum, for three benchmark masses of the lightest neutralino that are within the kinematically allow range $m_{\tilde{\chi}_1^0}<m_Z/2$: 20, 30, and 40 GeV.
In general we find the selection of $p_T^e>20$ GeV retains a large proportion of the events.
In the lower panel of Fig.~\ref{fig:distributions_n1} we show the distributions of the time delay for the same masses with $c\tau_{\tilde{\chi}_1^0}$ fixed at 1 m.
One easily observes that for the neutralinos with a larger mass, the time delay tends to be enhanced, allowing for better acceptance.
This is mainly due to the lowered speed of the heavier LLPs.
\begin{figure}[t]
	\includegraphics[width=0.49\textwidth]{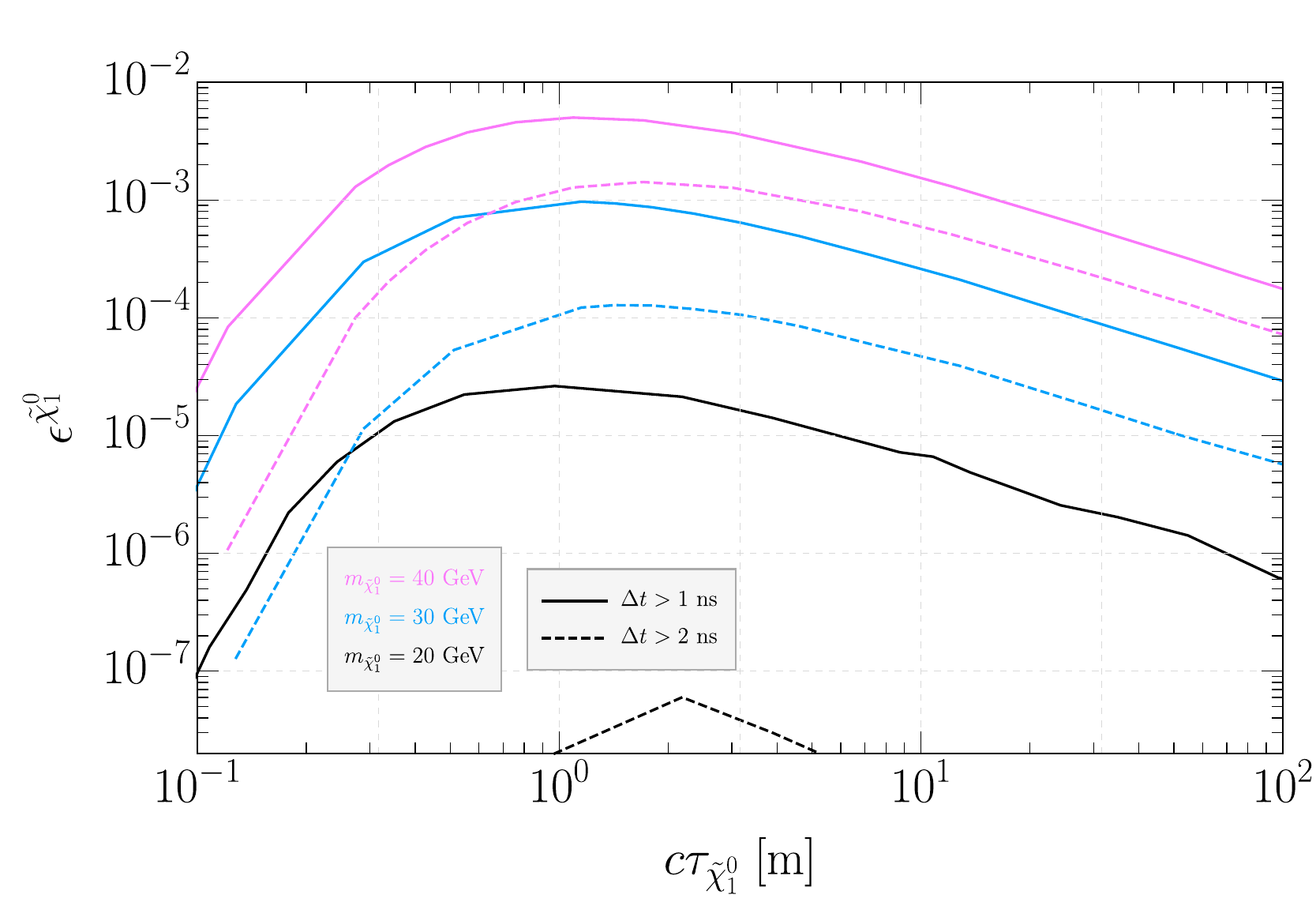}
	\caption{Acceptance $\epsilon^{\tilde{\chi}_1^0}$ as a function of $c\tau_{\tilde{\chi}_1^0}$, for three values of $m_{\tilde{\chi}_1^0}$.}
	\label{fig:accpvsctau_n1}
\end{figure}
The final acceptance rate $\epsilon^{\tilde{\chi}_1^0}$ is shown in Fig.~\ref{fig:accpvsctau_n1} as a function of $c\tau_{\tilde{\chi}_1^0}$ for three neutralino masses and for $\Delta t>1$ (solid lines) or 2 ns (dashed lines).
It is clear that the best acceptance rate is achieved at $c\tau_{\tilde{\chi}_1^0}\sim 1\text{ m}$ in general, and heavier neutralinos have a higher possibility to pass the event selection criteria.
For the larger masses, imposing a time-delay cut of 2 ns reduces the acceptance by a factor of about 5 at $c\tau_{\tilde{\chi}_1^0}\sim 1\text{ m}$ compared to $\Delta t > 1$ ns, while for $m_{\tilde{\chi}^0_1}=20$ GeV the reduction is much more severe.

\begin{figure}[t]
	\includegraphics[width=0.49\textwidth]{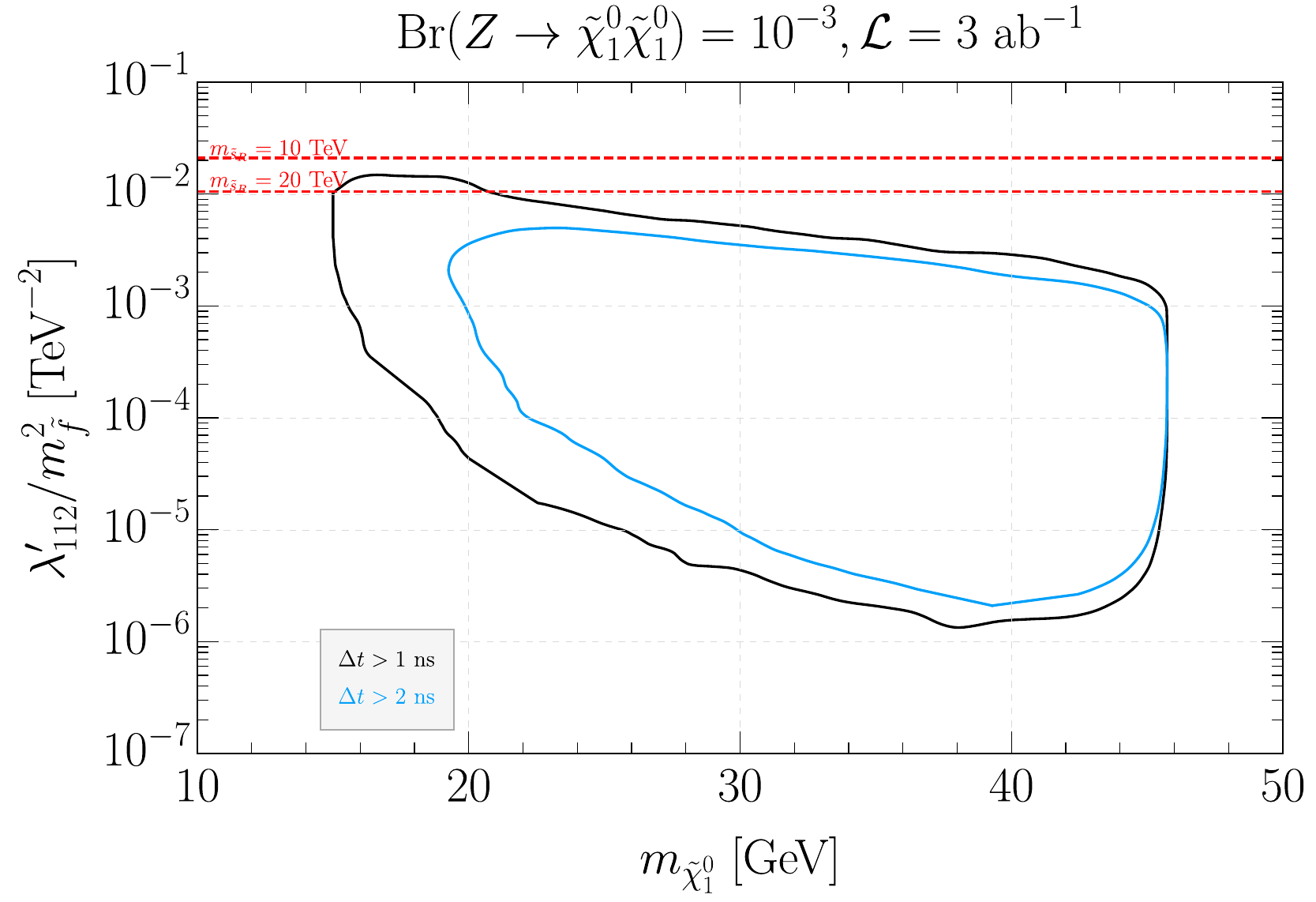}
	\includegraphics[width=0.49\textwidth]{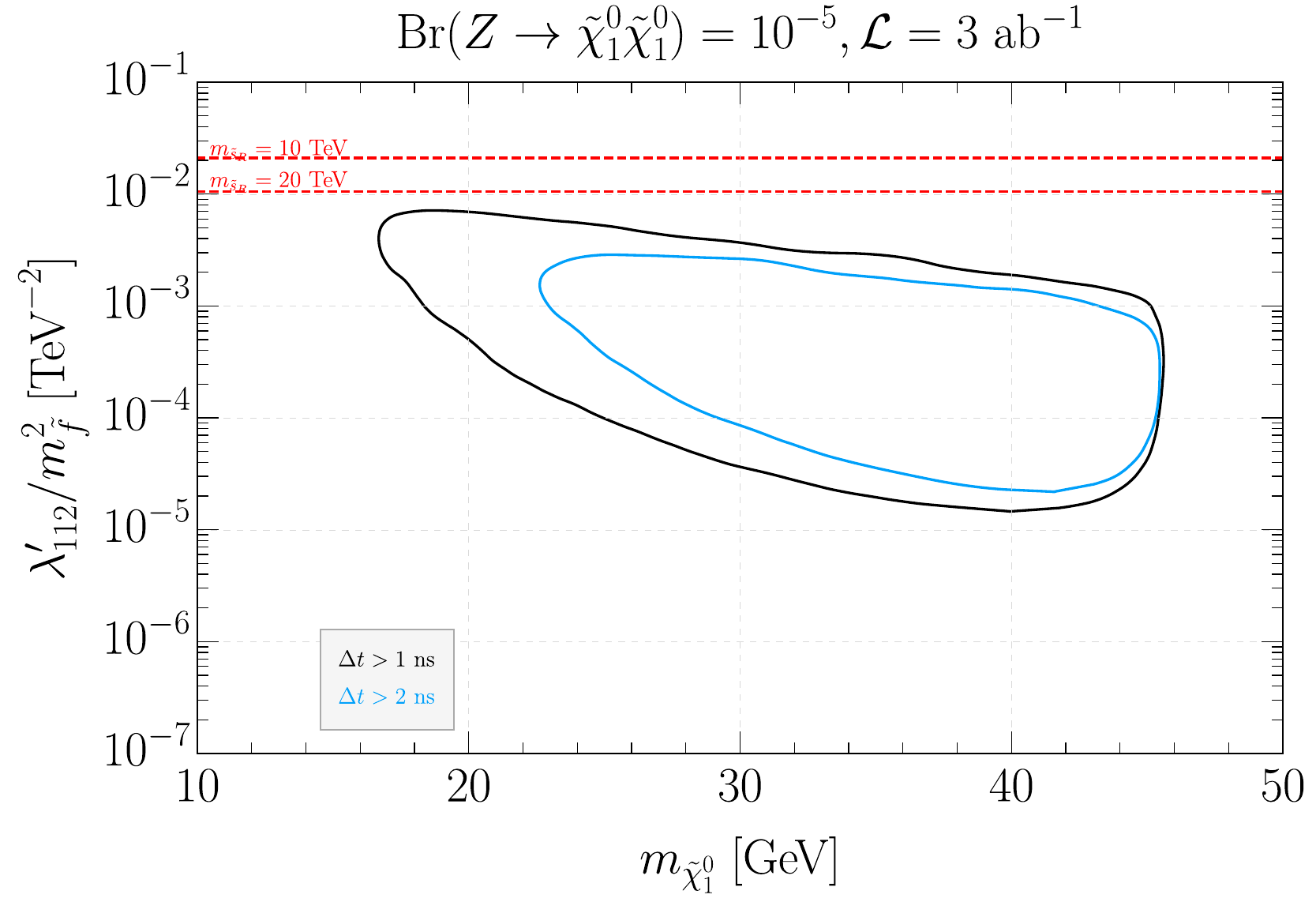}
\caption{$95\%$ C.L. exclusion limits in the $\lambda'_{112}/m^2_{\tilde{f}}$ vs. $m_{\tilde{\chi}_1^0}$ plane for Br($Z\to \tilde{\chi}_1^0\tilde{\chi}^0_1$)$=10^{-3}$ and $10^{-5}$.
The horizontal dashed lines correspond to the present limits with $m_{\tilde{s}_R}=10$ and 20 TeV (see Eq.~\ref{eqn:RPVbound}).	
}
	\label{fig:couplingvsmass_n1}
\end{figure}
In Fig.~\ref{fig:couplingvsmass_n1} we plot the sensitivity reaches in the
$\lambda'_{112}/m^2_{\tilde{f}}$ vs. $m_{\tilde{\chi}_1^0}$ plane for two benchmark values of Br($Z\to \tilde{\chi}_1^0\tilde{\chi}^0_1$)$=10^{-3}$ and $10^{-5}$.
The horizontal dashed lines are the present limits on $\lambda'_{112}/m^2_{\tilde{s}_R}$ for $m_{\tilde{s}_R}=10$ and 20 TeV (see Eq.~\eqref{eqn:RPVbound}).
In both plots, our search strategy is expected to probe the model parameter $\lambda'_{112}/m^2_{\tilde{f}}$ orders of magnitude stronger than the present bounds.
Imposing a more strict time-delay cut reduces the sensitivity reach the most in the small mass regime, while for $m_{\tilde{\chi}_1^0}$ close to the kinematic threshold, we observe relatively milder reduction in the limits.
The exclusion limits are bounded from low mass regime (left), mainly because of the faster speed of the LLP rendering the events less likely to pass the time-delay cut.
For too large or small values of $\lambda'_{112}/m^2_{\tilde{f}}$, the light neutralino tends to decay outside the fiducial volume.
Finally the kinematic constraint requires that $m_{\tilde{\chi}_1^0}<m_Z/2$.

A displaced-vertex (DV) search for the same theoretical scenario was proposed in Ref.~\cite{Wang:2019orr} for ATLAS with $3\text{ ab}^{-1}$ integrated luminosity.
The authors estimated the SM background from an ATLAS search \cite{Aad:2019xav} for a similar decay topology (Higgs decay to a pair of LLPs).
However, in Ref.~\cite{Wang:2019orr} no kinematic cuts were imposed on signal events and the whole ATLAS detector was taken as the fiducial volume, resulting in over-optimistic limits.
A more realistic DV search with $p_T$ cuts and a smaller fiducial volume (consisting of, for instance, only the inner tracker), would weaken the excluding potential especially for the small mass or coupling regimes.
Therefore, for fairness, we choose not to compare the sensitivity reach of the search strategy presented in this work directly with the results obtained in Ref.~\cite{Wang:2019orr}.

\begin{figure}[t]
	\includegraphics[width=0.49\textwidth]{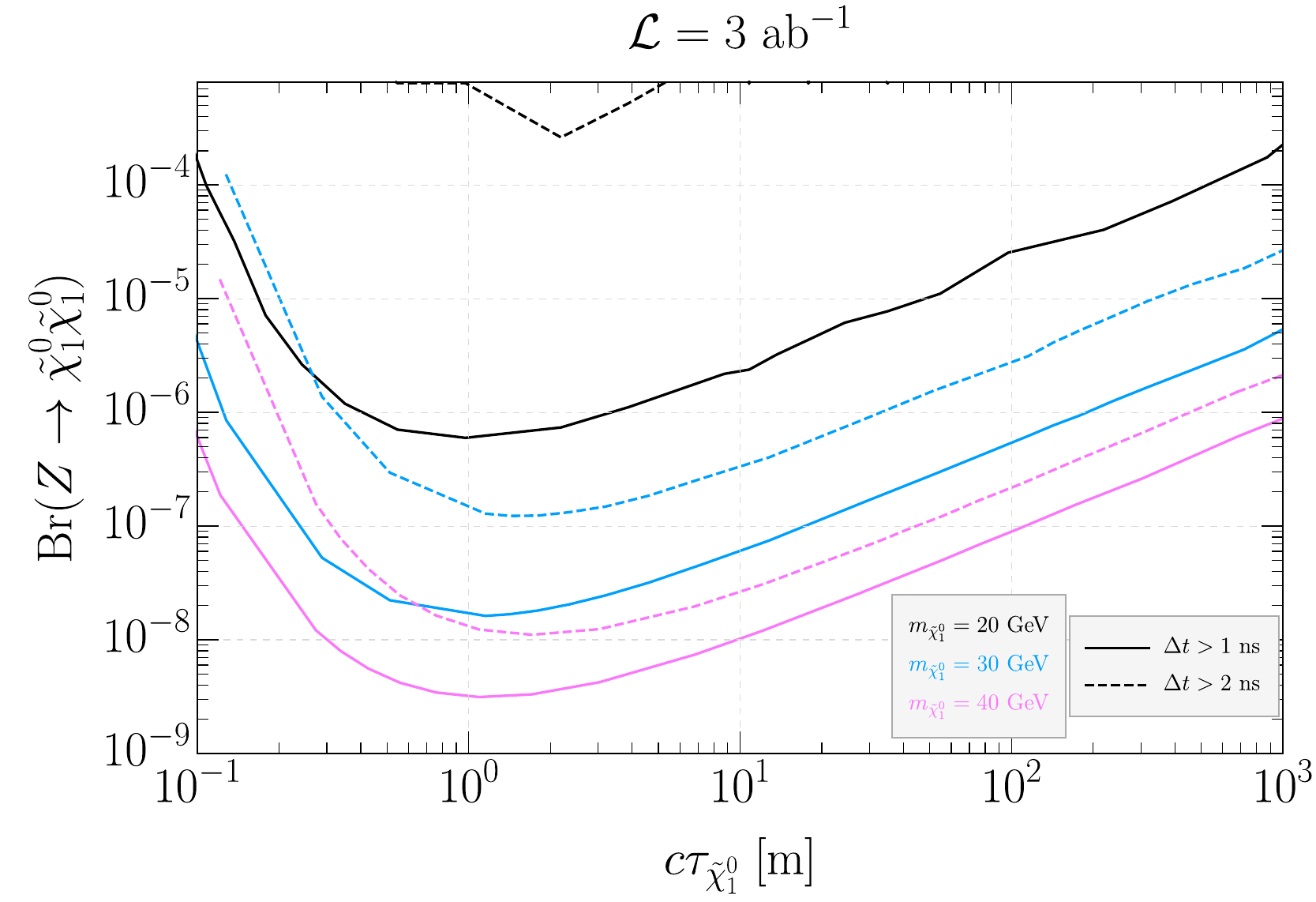}
	\caption{Sensitivity reach shown in the plane Br($Z\to \tilde{\chi}_1^0\tilde{\chi}_1^0$) vs. $c\tau_{\tilde{\chi}_1^0}$, for $m_{\tilde{\chi}_1^0}=20,30,40$ GeV.}
	\label{fig:brvvsctau_n1}
\end{figure}
We further present the sensitivity results in another fashion.
Fig.~\ref{fig:brvvsctau_n1} shows the projected exclusion limits on Br($Z\to \tilde{\chi}_1^0\tilde{\chi}^0_1$) for $m_{\tilde{\chi}_1^0}=20,30$, and 40 GeV, with varying $c\tau_{\tilde{\chi}_1^0}$ between 10 cm and 1 km.
We reach the following conclusions.
We find for heavier neutralinos we can probe smaller values of Br($Z\to \tilde{\chi}_1^0\tilde{\chi}^0_1$), and a more strict cut on the time-delay would result in weaker exclusion limits.
In particular, with an integrated luminosity of 3 ab$^{-1}$, we may probe Br($Z\to \tilde{\chi}_1^0\tilde{\chi}^0_1$) down to between $10^{-9}$ and $10^{-8}$ for the heavier neutralinos.

\subsection{The heavy neutral lepton scenario}\label{subsec:results2}

We proceed to discuss the numerical results for the HNL scenario.
As noted earlier we consider one Majorana HNL mixed only with the electron neutrino.

\begin{figure}[t]
	\includegraphics[width=0.49\textwidth]{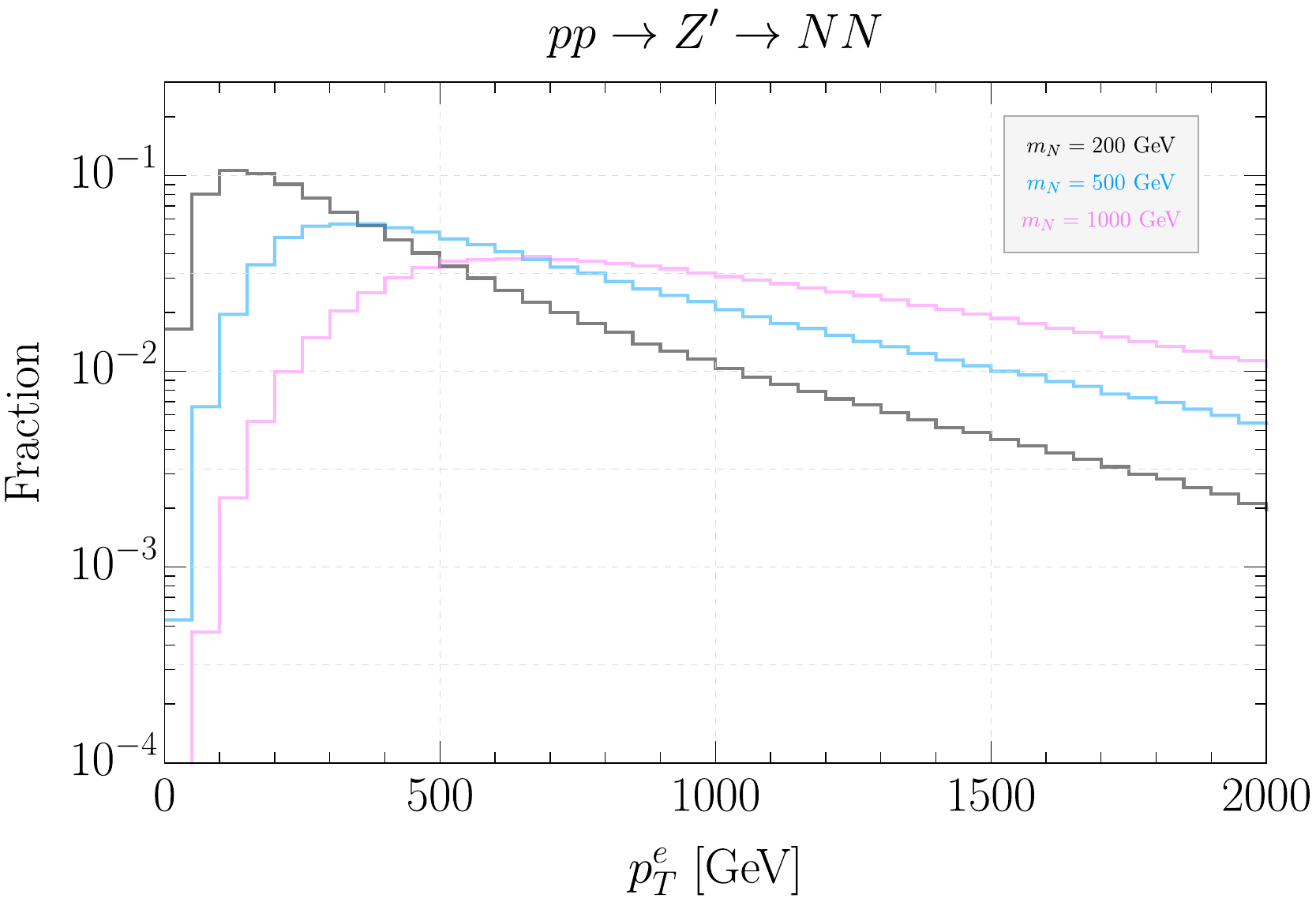}
	\includegraphics[width=0.49\textwidth]{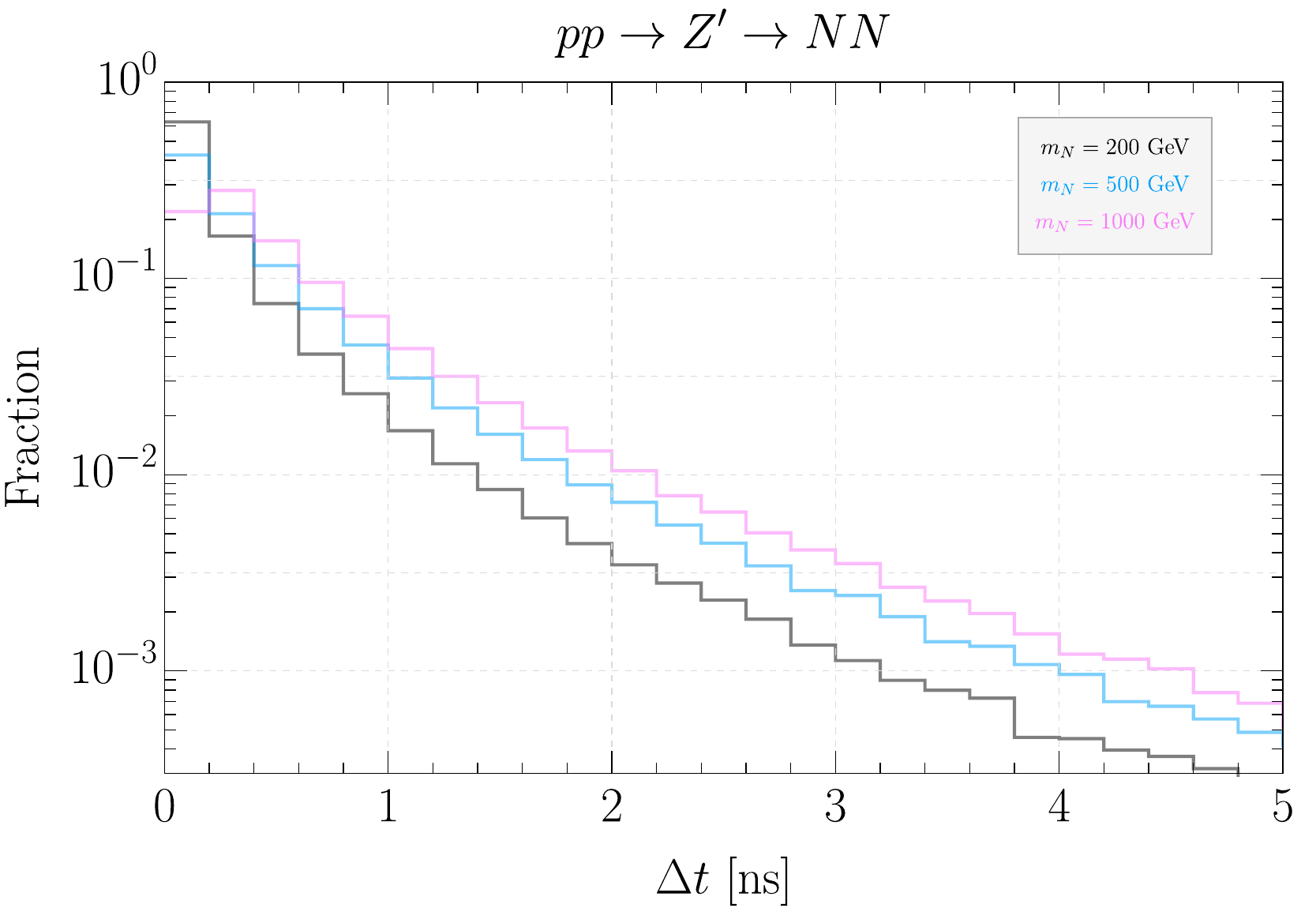}
	\caption{Distributions of $p_T^e$ and $\Delta t$ for HNLs. $c\tau_{N}$ is fixed at 1 m.}
	\label{fig:distributions_hnl}
\end{figure}
Since we assume a $Z'$ boson of mass 6 TeV, the kinematically allowed mass range of the HNL is much larger than that of the light neutralinos produced from the SM $Z$-boson decays. 
Fig.~\ref{fig:distributions_hnl} presents the distributions of the leading electron transverse momentum $p_T^e$ and the time delay $\Delta t$, for HNL masses of 200, 500, and 1000 GeV with $c\tau_{N}$ fixed at 1 m.
The upper panel shows clearly that the $p_T^e> 20$ GeV cut would have an unsubstantial effect on the signal events.
In the lower plot, we find that given the relatively large mass of the HNLs, a larger fraction of the HNLs are expected to have a time-delay larger than 1 or 2 ns, compared to the light neutralinos shown in Fig.~\ref{fig:distributions_n1}.
We note that the two $U(1)$ extension alternatives differ, phenomenologically speaking, only in $\sigma^N$, and share the same kinematics.
\begin{figure}[t]
	\includegraphics[width=0.49\textwidth]{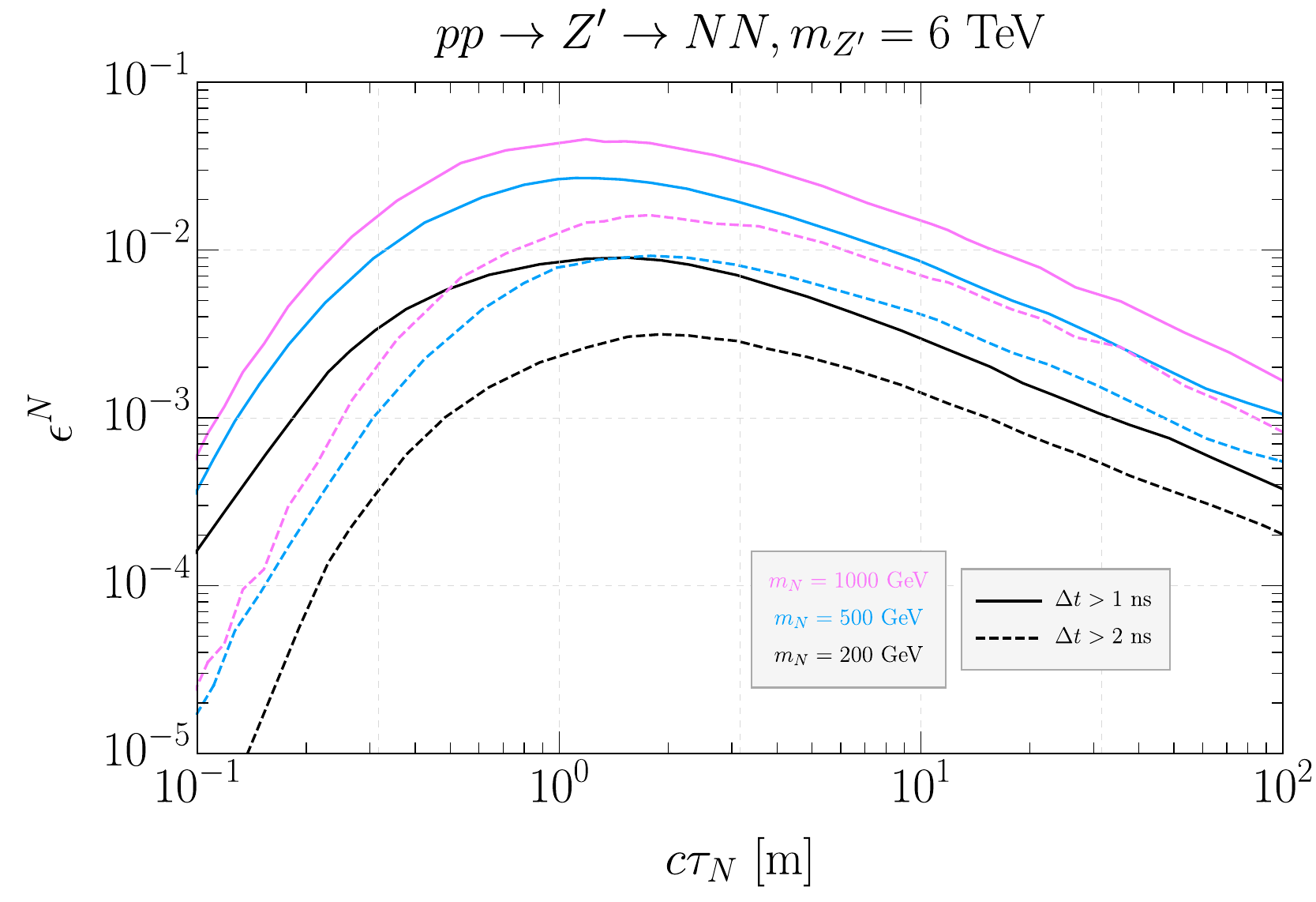}
	\caption{The acceptance rate $\epsilon^N$ as a function of $c\tau_{N}$, for three values of $m_{N}$.}
	\label{fig:accpvsctau_hnl}
\end{figure}
The final acceptance is given in Fig.~\ref{fig:accpvsctau_hnl} as a function of $c\tau_{N}$ for $m_N = 200, 500,$ and 1000 GeV, where two choices of the $\Delta t$ cut are taken.
Since the kinematics of the HNLs in $U(1)_{X}$ and $U(1)_{B-L}$ are the same, it suffices to show the acceptance rates for only one model.
Similar to the neutralino scenario, here the maximal acceptance is also achieved at proper decay lengths around 1 m, and heavier HNLs have a higher chance of passing all the event cuts.
This conclusion was also drawn in Ref.~\cite{Mason:2019okp} where a similar topology (SM Higgs decaying to a pair of HNLs) with the same timing strategy was studied.

\begin{figure}[t]
	\includegraphics[width=0.49\textwidth]{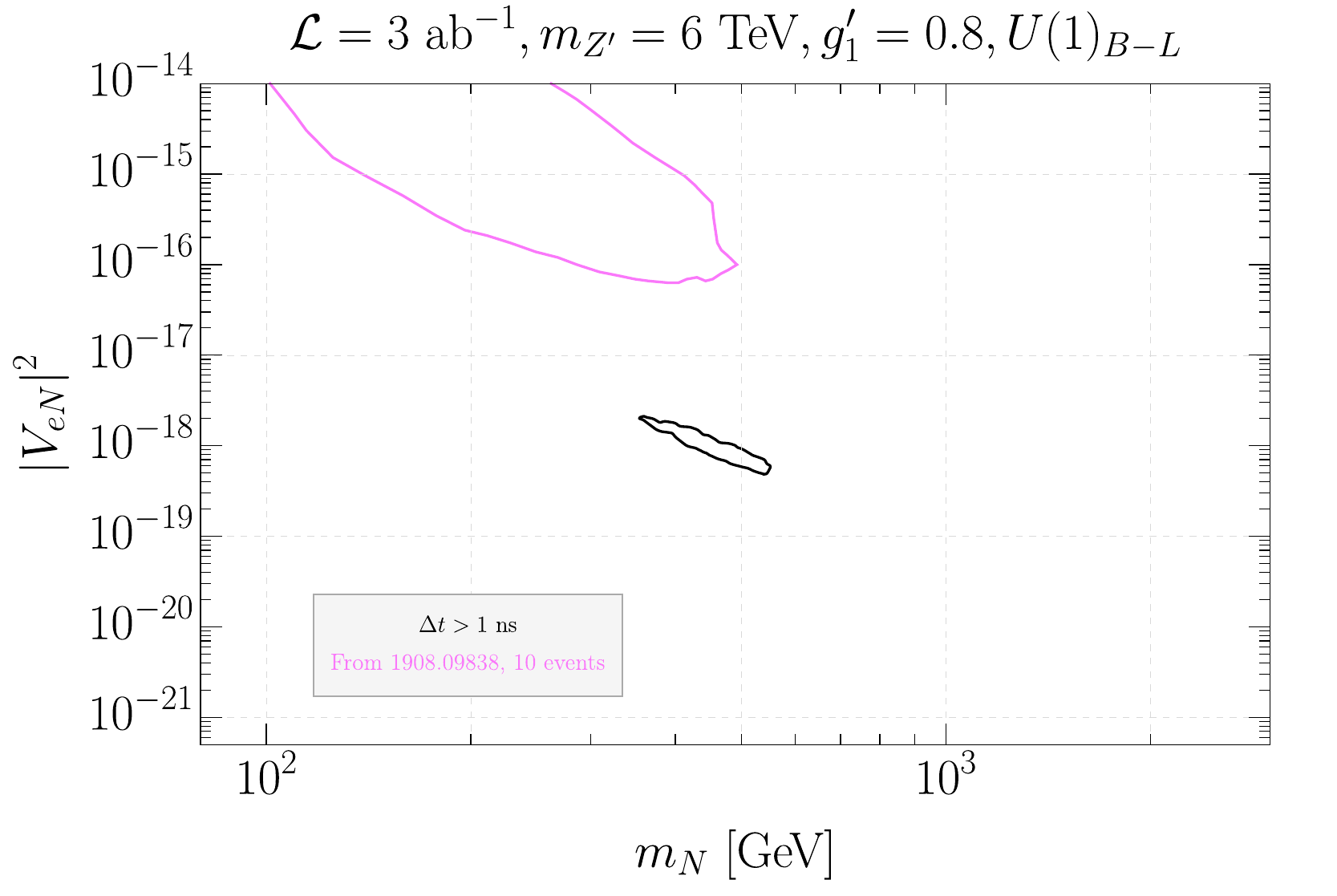}
	\includegraphics[width=0.49\textwidth]{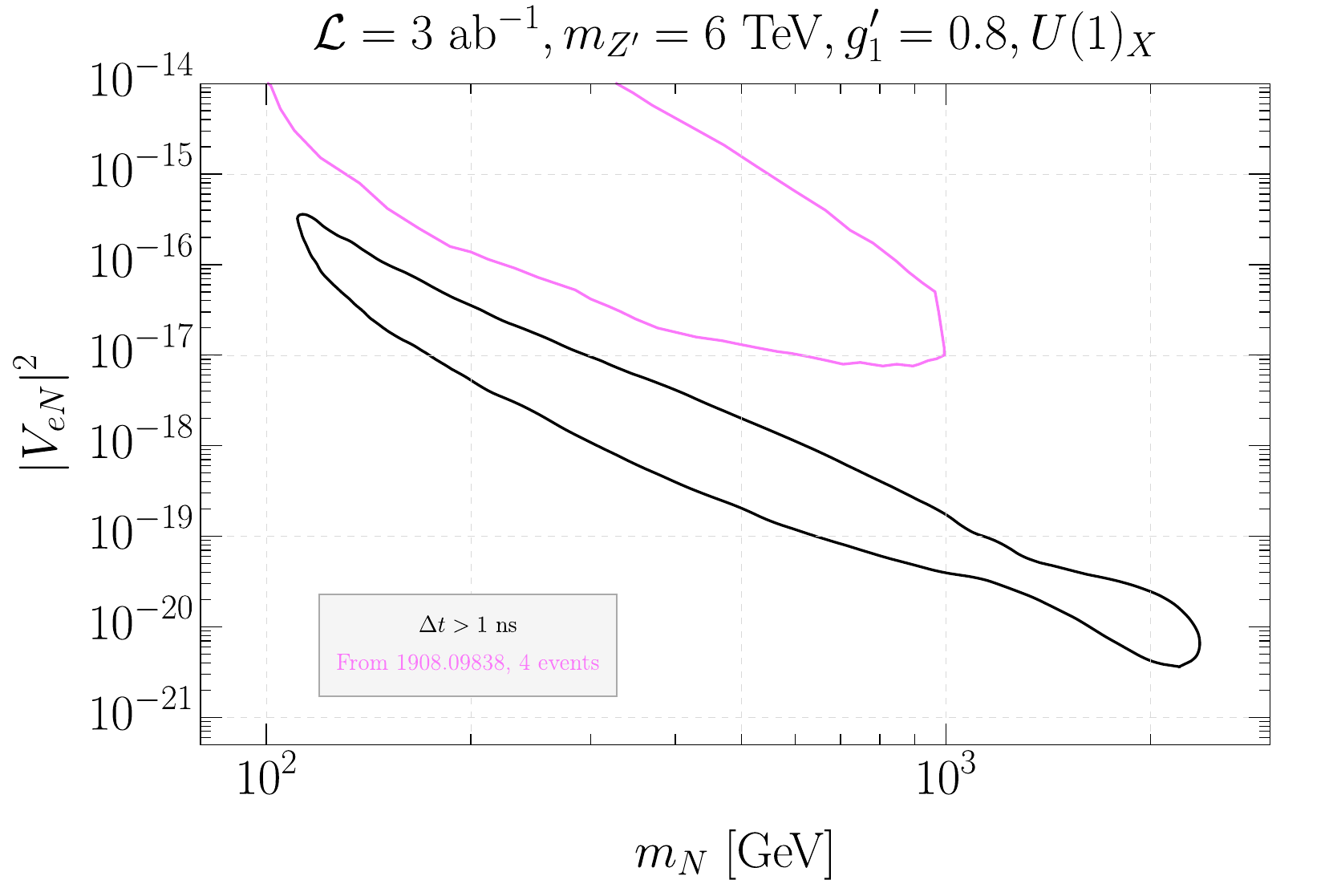}
	\caption{The 95\% C.L. sensitivity reaches in the
          $|V_{eN}|^2$ vs. $m_{N}$ plane for both $U(1)_{B-L}$ and $U(1)_X$ models.
	The pink curves are extracted from Ref.~\cite{Chiang:2019ajm}.
	}
	\label{fig:couplingvsmass_hnl}
\end{figure}
For sensitivity plots, we first show in Fig.~\ref{fig:couplingvsmass_hnl} the reaches in the $|V_{eN}|^2$ versus $m_N$ plane, for both $U(1)_{B-L}$ and $U(1)_X$ models.
The projected sensitivity limits from a DV search \cite{Chiang:2019ajm} are shown together in the same plots.
We find that with a stringent cut of $\Delta t>2$ ns, no sensitivity can be achieved, while requiring $\Delta t>1$ ns allows to probe certain parts of the parameter space which are inaccessible by the DV search \cite{Chiang:2019ajm}.
The $U(1)_{B-L}$ scenario, because of its relatively small scattering cross section, is expected to achieve very limited constraining power in the $m_N-|V_{eN}|^2$ plane, while in the $U(1)_X$ case a rather long band of the parameter space can be probed.
These plots in Fig.~\ref{fig:couplingvsmass_hnl} exemplify clearly the complementarity of the timing-trigger-based search to the other strategies of LLP searches.

\begin{figure}[t]
	\includegraphics[width=0.49\textwidth]{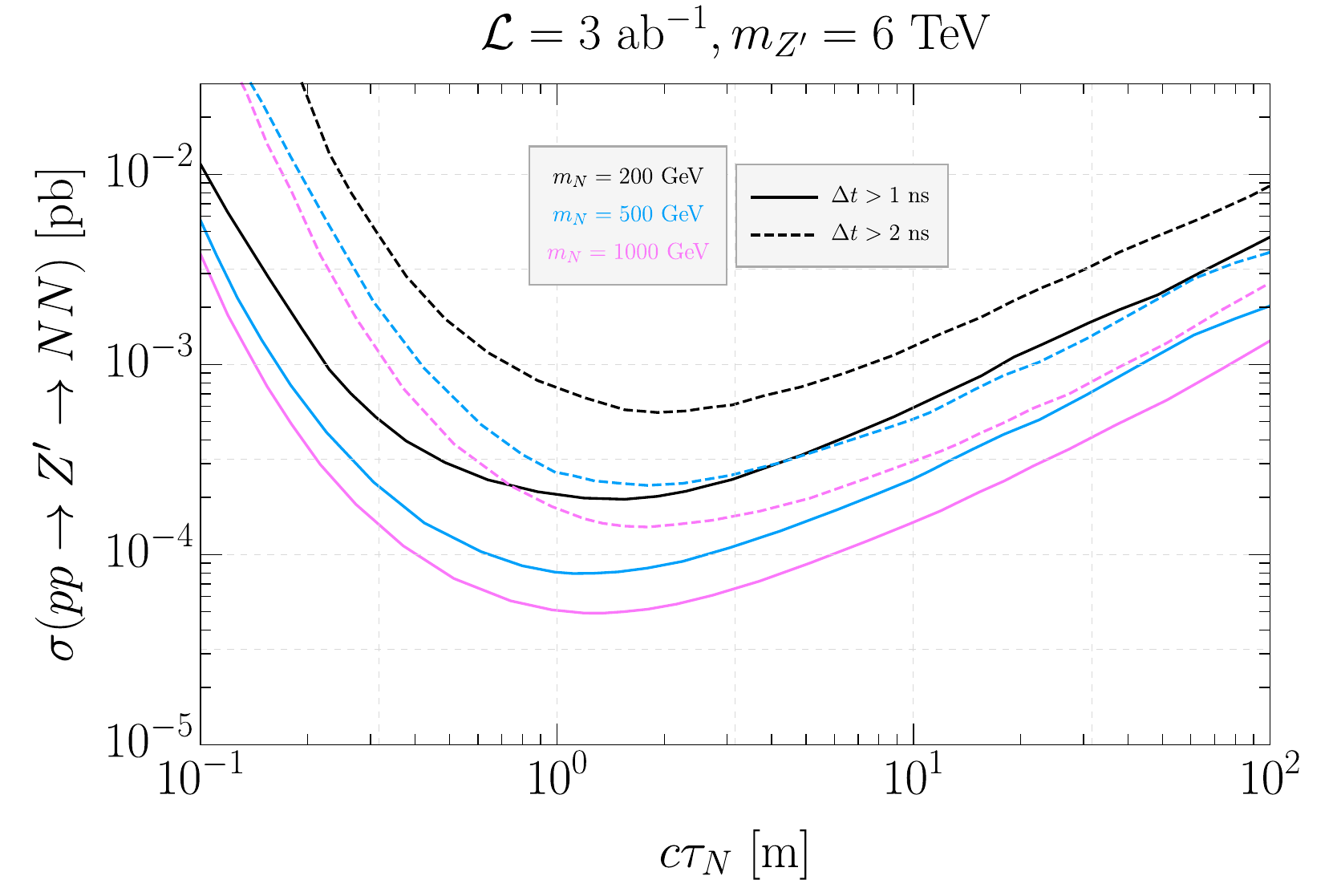}
	\caption{95\% C.L. xxclusion limits on $\sigma^N$ when varying $c\tau_{N}$, for $m_{N}=200,500$, and 1000 GeV.}
	\label{fig:sigmaNvsctau_hnl}
\end{figure}
Finally, we obtain the exclusion limits in the $c\tau_{N}-\sigma^N$ plane for $m_N=200,500,$ and 1000 GeV, as shown in Fig.~\ref{fig:sigmaNvsctau_hnl}.
We observe that stronger limits are expected for heavier HNLs, unless approaching the kinematic threshold of 3000 GeV which is not shown here.
For instance, for $m_N=1000$ GeV, with a $\Delta t > 1$ ns cut, limits on $\sigma^N$ can be achieved at as low as $5\times 10^{-5}$ pb when $c\tau_{N}\simeq 1\text{ m}$. 
These results can be used to constrain other models with similar kinematics and the same scattering and decay topologies, i.e., a new 6-TeV particle produced from the $pp$ collisions and decaying to a pair of LLPs which subsequently decay into $ejj$.
To better facilitate this purpose, we list the values of Br$(N \to ejj)$ for $m_N =200, 500,$ and 1000 GeV: $28.6\%,23.4\%,\text{ and } 22.7\%$, computed in this study.

\subsection{Discussion}\label{sec:discuttion}

The two benchmark models, though sharing similar scattering and decay topologies as well as collider signatures, still differ in certain aspects such as the kinematics essentially because of the different $s$-channel resonance masses.
In the $U(1)$-extension models, the $Z'$ has to be heavy to be consistent with the previous experimental results.
Here we assume it has a mass of 6 TeV, which is almost two orders of magnitude heavier than its counterpart, the SM $Z$-boson, in the light neutralino scenario.
This allows for probing much heavier LLPs which would also decay to electrons with a larger $p_T^e$.
More concretely, in this work, kinematically allowed range of $m_N$ is up to 3 TeV, in comparison with $\sim 45$ GeV for the light neutralinos.
This makes it possible to probe smaller values of the feeble couplings to the SM particles in the HNL scenario than in the neutralino one.
Moreover, in general, heavier LLPs travel more slowly or even more non-relativistically, improving significantly the time-delay search acceptance.

We should also provide some further comments on the lepton flavors.
In this work, we have focused on the electron case, i.e. the HNL mixes only with $\nu_e$ and the $L_i \cdot Q_j\bar{D}_k$ operator has $i=1$.
However, it is also possible to have a muon or even a tau lepton in the final state, for HNLs mixed with $\nu_{\mu/\tau}$ and $L_{2/3} \cdot Q_j \bar{D}_k$ operators.
In principle, since the timing layers are based on ionization processes,
as long as a charged particle hits them, timing information can be stored.
For the muon final state, since muons travel relativistically, we expect
the sensitivity results should not change qualitatively from those for
the electron final state, except for some minor discrepancies resulting from the muon masses.
However, a final-state tau lepton decays fast into hadrons dominantly.
Consequently, the triggering will be more difficult and complicated, and the corresponding exclusion power should be weakened.

Furthermore, we consider here the HNL to be of Majorana nature.
It is possible that the HNLs are Dirac fermions.
In this case, the decay width of the HNL should halve for the same mass and mixing, thus reducing the acceptance and hence weakening the final sensitivity reach in $|V_{eN}|^2$ by approximately a factor of 2 in the large decay length limit (with small $m_N$ or $|V_{eN}|^2$).

In our study we have assumed a somewhat optimistic efficiency, 100\%, for the timing trigger.
Once the realistic efficiency is known\footnote{In Refs.~\cite{Liu:2018wte,Mason:2019okp}, a timing-trigger efficiency of 50\% was assumed as a benchmark value.}, the sensitivity reach should weaken accordingly.

Finally, as the upper plots of Fig.~\ref{fig:distributions_n1} and Fig.~\ref{fig:distributions_hnl} show, varying the $p_T^e$ threshold, say, between 10 and 30 GeV, should not affect our exclusion limits significantly, especially for the heavier HNLs in the $Z'$ scenario.

\section{Conclusions}\label{sec:conclusions}

At various LHC experiments, future upgrades in the timing detectors have been proposed primarily for the purpose of reducing pileup in the HL-LHC phase.
However, the precision timing information from such setup can also
be used to enhance the discovery potential for long-lived particles at colliders.
In this work, we focus on the CMS minimum-ionizing-particle timing detector, and follow existing literature to propose a timing-based search for the two types of neutral LLPs in similar channels.

We have investigated the sensitivity reaches for long-lived light neutralinos and heavy neutral leptons in the context of two well-motivated theoretical models, by searching for a hard and time-delayed electron from neutral currents.

Light neutralinos are still allowed in the R-parity-violating   supersymmetry because of the decay of the lightest neutralino into SM particles.
  We consider the SM $Z$-boson decay into a pair of the lightest neutralinos, which become long-lived for small mass as well as tiny
RPV couplings ($\lambda'_{112}/m^2_{\tilde{f}}$ as considered in this work), assuming degenerate sfermion masses.
For the HNLs we study two different $U(1)$ extensions of the SM, where a heavy new gauge boson $Z'$ can be produced on-shell at the LHC and decays
to a pair of $N$'s, which are long-lived for small mixings with the active neutrinos.
For both scenarios, we focus on the final state $ejj$ from the LLP decays.
The main background sources are same-vertex hard collisions and pileup events.
These have been estimated in Ref.~\cite{Mason:2019okp} to be negligible, if one requires $\Delta t \gtrsim 1$ ns.

We simulate the production and decay processes for various LLP masses and couplings to take into account their effects on the production rate, kinematics, etc. Considering the geometry and precision of the proposed timing layer, the final state data are analyzed in order to extract the $95\%$ C.L. exclusion limits in the model parameter space.
We present the results in terms of isocurves in both the
$\lambda'_{112}/m^2_{\tilde{f}}$ vs. $m_{\tilde{\chi}_1^0}$ and Br$(Z\to \tilde{\chi}_1^0\tilde{\chi}_1^0 )$ vs. $c\tau_{\tilde{\chi}_1^0}$ planes for the light neutralino scenario, and in both the $|V_{eN}|^2$ vs. $m_N$ and $\sigma^N$ vs. $c\tau_N$ planes for the heavy neutral lepton scenario.
The results indicate that our search strategy would be able to probe complementary parameter regions, compared to traditional strategies such as those based on displaced vertex.
In particular, we find the acceptance rate tends to get enhanced for heavier LLPs.
Moreover, our ``model-independent'' limits in the Br$(Z\to \tilde{\chi}_1^0\tilde{\chi}_1^0 )$ vs. $c\tau_{\tilde{\chi}_1^0}$ and $\sigma^N$ vs. $c\tau_N$ planes can be used to constrain other theoretical models with similar kinematics, scattering topology, and decay products.

In conclusion, we have demonstrated that a search strategy based on a timing trigger has the potential to probe parameter space that is complementary to other types of LLP searches.
If we can combine the displaced-vertex search with the time-delay search, we can substantially improve the coverage of the parameter space. 
In particular, compared to charged LLPs, neutral LLPs are more elusive, and searches for them usually require more sophisticated or smarter methods, e.g. the timing trigger strategy discussed in this work.
We expect that more uncharted territories in the parameter space of other theoretical models with LLPs can be explored with the proposed upgrades in the timing detectors at the LHC and this novel type of search strategy.

\vspace{0.5cm}
\textit{Note added after publication:} we thank Matthew Strassler for pointing out the fact that the time-delay technique for detecting LLPs, often in the context of delayed photons or leptons in Gauge Mediated Supersymmetry Breaking scenarios, was originally brought up in the early 2000's~\cite{Toback:2004xd,Prieur:1019876,Meade:2010ji,Hong:2012sp}, and has been applied in experimental searches at \textit{e.g.} CDF~\cite{CDF:2007sit,CDF:2008xrp,CDF:2013sjb}, ATLAS~\cite{ATLAS:2014kbb}, and CMS~\cite{CMS-PAS-EXO-12-035,CMS:2012bbi,CMS:2019zxa}.

\section*{Acknowledgment}
We thank Giovanna Cottin for sharing the UFO model files, and thank Van Que Tran and Shin-Shan Eiko Yu for useful discussions on the timing detector.
Z.S.W. is supported by the Ministry of Science and Technology (MoST) of Taiwan with grant number MoST-109-2811-M-007-509.
K.C. was supported by MoST with grant numbers 
MoST-107-2112-M-007 -029 -MY3 and MOST-110-2112-M-007-017-MY3.
K.W. is supported by the National Natural Science Foundation of China under grant no.~11905162, 
the Excellent Young Talents Program of the Wuhan University of Technology under grant no.~40122102, the research program of the Wuhan University of Technology under grant no.~3120620265.

\bibliography{bib}
  
\end{document}